\documentclass[aps,prl,letterpaper,superscriptaddress,amsmath,amssymb,twocolumn]{revtex4}

\usepackage[usenames,dvipsnames]{color}
\usepackage[colorlinks,citecolor=blue]{hyperref}
\usepackage{amsmath}
\usepackage{bbm}
\usepackage{amsfonts}
\usepackage{amssymb}
\usepackage{latexsym}
\usepackage{graphicx}
\usepackage[english]{babel}
\usepackage{multirow}
\usepackage{float}
\usepackage{url}
\usepackage{slashed}
\usepackage{xcolor} 
\usepackage{graphicx} %use graph format
\usepackage{epstopdf}

\newcommand{\be}{\begin{equation}}
\newcommand{\ee}{\end{equation}}
\newcommand{\ba}{\begin{array}}
	\newcommand{\ea}{\end{array}}
\newcommand{\bea}{\begin{eqnarray}}
\newcommand{\eea}{\end{eqnarray}}

\newcommand{\eesesea}{$e^+ e^- \to \slashed{e}^+  \slashed{e}^- \gamma$\,}

\newcommand{\eevva}{$e^+ e^- \to \nu\bar \nu \gamma$\,}

\newcommand{\GeV}{\,{\rm GeV}}

\newcommand{\nn}{\nonumber}

\newcommand{\mzp}{m_{Z^\prime}}
\newcommand{\lmt}{L_\mu-L_\tau}
\newcommand{\zp}{Z^\prime}
\newcommand{\epem}{e^+e^-}

\usepackage{ulem,fancyvrb}
\usepackage{xcolor}

\begin{document}
%\title{Probing  the $L_\mu-L_\tau$ gauge boson  at electron colliders via monophoton searches} 
\title{Probing  the $L_\mu-L_\tau$ gauge boson  at electron colliders}

\author{Yu Zhang} 
\affiliation{Institutes of Physical Science and Information Technology,
	Anhui University, Hefei 230601, China}
\affiliation{School of Physics and Materials Science, Anhui University, Hefei 230601,China }

\author{Zhuo Yu}
\affiliation{Institutes of Physical Science and Information Technology,
	Anhui University, Hefei 230601, China}
\author{Qiang Yang}
\affiliation{Zhejiang Institute of Modern Physics, Department of Physics, Zhejiang University, Hangzhou, 310027, China}

\author{Mao Song}
\affiliation{School of Physics and Materials Science, Anhui University, Hefei 230601,China }

\author{Gang Li}\email{lig2008@mail.ustc.edu.cn}
\affiliation{School of Physics and Materials Science, Anhui University, Hefei 230601,China }

\author{Ran Ding} 
\affiliation{School of Physics and Materials Science, Anhui University, Hefei 230601,China }

\begin{abstract}
We investigate the minimal  $U(1)_{\lmt}$ model  with extra heavy vector-like 
leptons or charged scalars.
By studying the kinetic mixing between $U(1)_{\lmt}$ gauge boson $\zp$ and standard model photon,
which is absent at tree level and will arise at one loop level due to $\mu$, $\tau$ and new heavy charged
leptons or scalars, the interesting behavior is shown.
It can provide possibility for visible signatures  of new heavy particles.
We propose to search for  $\zp$ at electron collider experiments, such as Belle II, BESIII and future Super Tau Charm Factory (STCF), using the monophoton final state. The parameter space of $\zp$ is probed, and 
scanned by its gauge coupling constant $g_{\zp}$ and mass $\mzp$.
We find that electron colliders have sensitivity to the previously unexplored parameter space for $\zp$ with MeV-GeV mass.
Future STCF  experiments with $\sqrt s=2-7$ GeV can exclude the anomalous muon magnetic moment favored area when $\mzp<5$ GeV with the luminosity of 30 ab$^{-1}$.
For $\mzp < 2m_\mu$, $g_{\zp}$ can be down to $4.2\times 10^{-5}$ at 2 GeV STCF.

\end{abstract}

	\maketitle

\section{Introduction}

The standard model (SM) of particle physics is a successful and highlypredictive theory of
fundamental particles and interactions, but fails to explain many phenomena, including
neutrino mass, baryon asymmetry of the universe, presence of dark matter (DM) and dark
energy, among others. It implies that SM is only a low-energy approximation of the more
fundamental theory; extensions of SM are strongly required.

Among various extended scenarios beyond SM, new U(1) gauge symmetries
are of particular 
interest since this is one of the minimal extensions of the SM.
In particular, the $U(1)_{\lmt}$ model 
\cite{Foot:1990mn,He:1990pn,He:1991qd}, with a $U(1)_{\lmt}$ extension of SM, 
gauges the difference of the leptonic muon and tau number and induces a new vector boson $\zp$. 
This model has gained a lot of attention, since it can be free from gauge anomaly without any extension of particel content.
Moreover, it is potentially able to address important open issues in particle physics,
such as the discrepancy in moun anomalous magnetic moment $(g-2)_\mu$
\cite{Bennett:2006fi,Baek:2001kca,Ma:2001md,Altmannshofer:2016brv}, $B$ decay anomalies
\cite{Aaij:2013qta,Aaij:2014ora,Altmannshofer:2016jzy,Crivellin:2015mga,Baek:2017sew,Ko:2017yrd} and recent  anomalous  excess  in $K_L\to\pi^0+{\rm INV}$
\cite{Jho:2020jsa}.
Besides, the $U(1)_{\lmt}$ model has also been discussed in lepton-flavor-violating decay of the Higgs boson\cite{Crivellin:2015mga,Altmannshofer:2016oaq}, the neutrino masses and mixing
\cite{Ma:2001md,Heeck:2011wj,Baek:2015mna,Biswas:2016yan}
, and dark matter \cite{Biswas:2016yan,Altmannshofer:2016jzy,Patra:2016shz,Biswas:2016yjr,Biswas:2017ait, Arcadi:2018tly,Kamada:2018zxi,Foldenauer:2018zrz,Cai:2018imb,Han:2019diw}.

Since $\zp$ can directly couple to muon, related searches for $\zp$ have been performed
with the production of $\mu^+\mu^- \zp$ at collider experiments,
including BaBar \cite{TheBABAR:2016rlg} and Belle II \cite{Adachi:2019otg} at electron colliders
and CMS \cite{Sirunyan:2018nnz} at hadron collider. 
Subsequently, $\zp$ decaying to muon-pair is considered at BaBar and CMS experiments,
and invisible decay of $\zp$ is considered at Belle II.
Phenomenally, Ref. \cite{Jho:2019cxq} investegated the sensitivity on $\zp$ at Belle II with 
the planned target luminosity of 50 ab$^{-1}$ in the channel of
$e^+e^-\to \mu^+\mu^- \zp, \zp\to {\rm INV}$;
Refs. \cite{Kaneta:2016uyt,Araki:2017wyg, Chen:2017cic,Banerjee:2018mnw} proposed the search for $\zp$ at Belle II using the monophoton process
$e^+e^-\to \gamma \zp, \zp\to {\rm invisible}$,
which depends on the kinetic mixing between the SM photon and $\zp$.

In this work, we investigate the $\gamma-\zp$ kinetic mixing in the
minimal $U(1)_{\lmt}$ with extra heavy vector-like leptons 
or charged scalars.
Then we propose to search for $\lmt$ gauge boson  $\zp$ at electron collider experiments, such as Belle II, BESIII and future Super Tau Charm Factory (STCF), using the monophoton final state.
Belle II is an asymmetric detector and located at SuperKEKB  which collides 7 GeV 
electrons with 4 GeV positrons. 
SuperKEKB has a  largest  instantaneous luminosity  of $8\times 10^{35}$ cm$^{-2}$ s$^{-1}$ \cite{Kou:2018nap}.
The ambitious goal of SuperKEKB  is to accumulate an integrated luminosity of 50 $\rm{ab}^{-1}$ with 8-year data takings \cite{Kou:2018nap}.
The BESIII detector is  symmetric  and operated on the BEPCII 
with the beam energy ranging from 1.0 GeV to 2.3 GeV 
and a peak luminosity of $10^{33}$ cm$^{-2}$ s$^{-1}$ \cite{Asner:2008nq}. 
STCF is a proposed symmetric detector 
experiment which collides electron with positron in the range of  center-of-mass energies from 2.0 to 7.0 GeV
with the peak luminosity 
${\cal O}(10^{35})$ cm$^{-2}$ s$^{-1}$ 
at 4 GeV \cite{Peng:2019,Luo:2018njj,Bondar:2013cja}. 

The rest paper is organized as follows: First, we introduce the $U(1)_{\lmt}$ 
models and discuss their phenomenological features.
Then, we calculate the cross sections of the signal and the backgrounds and analysis to improve the significance by appropriate event cuts at three different electron colliders operated 
at the GeV scale: BelleII, BESIII and STCF.
The sensitivities for $\zp$ at these experiments are also investigated.
Finally, a short summary and discussions are given.

\section{The $U(1)_{\lmt}$ models }
	
\subsection{The minimal $U(1)_{\lmt}$ model}
We extend the SM with a new $U(1)$ gauge symmetry, $U(1)_{\lmt}$, where leptons of the second and third  generation
couple to the additional $U(1)_{\lmt}$ gauge boson $Z^\prime$ with equal and opposite charge. 
The new leptonic gauge interactions can be given as
\begin{equation}
%\mathcal{L}_{\rm int}^{\rm min} 
\mathcal{L}_{\rm int}
= g_{\zp} \left(\bar{\mu} \gamma^{\mu} \mu-\bar{\tau} \gamma^{\mu} \tau+\bar{\nu}_{\mu} \gamma^{\mu} P_L\nu_{\mu}-\bar{\nu}_{\tau} \gamma^{\mu} P_L \nu_{\tau }\right) \zp_\mu,
\end{equation}
where  $g_{Z^\prime}$ is gauge coupling constant.

In the minimal $U(1)_{\lmt}$ model, the kinetic mixing between the $\zp$ and 
photon is absent at the tree level. 
Nevertheless, because $\mu$ and $\tau$ 
are both charged under the electromagnetic $U(1)$ and $U(1)_{\lmt}$,  
there exists an unavoidable kinetic mixing at one loop level,
which can appear as \cite{Araki:2017wyg}

\begin{align}
\varepsilon^{\rm min}(q^2)=&\Pi(q^{2})
=
\begin{minipage}{4cm}
\unitlength=1cm
\begin{picture}(3.5,1.5)
\put(0,0){\includegraphics[width=3.5cm]{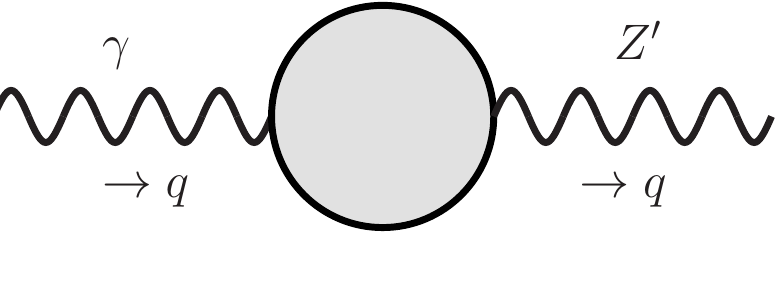}}
\end{picture}
\end{minipage}
\nonumber
\\
%%%%%
=&
\begin{minipage}{4cm}
\unitlength=1cm
\begin{picture}(3.5,1.5)
\put(0,0.05){\includegraphics[width=3.5cm]{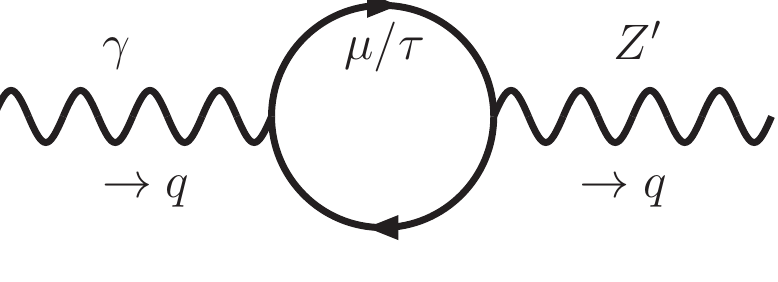}}
\end{picture}
\end{minipage}
\nonumber
\\
%%%%%
=&
\frac{8 eg_{\zp}^{}}{(4\pi)^2}
\int^1_0 x (1-x) {\rm ln}\frac{m_\tau^{2} - x(1-x)q^2}
{m_\mu^{2} - x(1-x)q^2} ~dx.
\label{eq:mixing-minimal}
\end{align}
Here $e$ is the electromagnetic charge, $m_\tau$ and $m_\mu$ are the masses of tau and muon leptons, $q$ is 
the transferred momentum.

For large momentum transfer $q^2\gg m_\tau^2$, this mixing is power suppressed by $1/q^2$, whereas for low momentum 
transfer $q^2\sim 0\ll m_\mu^2$, the mixing tends to be a constant 
\be
\varepsilon^{\rm min}(0)=\Pi(0)=\frac{e g_{Z^\prime}}{6\pi^2}
\ln \frac{m_\tau}{m_\mu},
\ee
which seems like the dark photon model.
	
\subsection{The $U(1)_{\lmt}$ model with extra heavy vector-like leptons}
We add two extra singlet vectorlike leptons ($L_1,\ L_2$) 
in the $U(1)_{\lmt}$ extension of the SM,
which are charged under $U(1)_{\lmt}$ opposite in sign 
similar as the $\mu$ and $\tau$, and have electric charge of $e$ \cite{Chen:2017cic}.
Since we mainly focus on the gauge kinetic mixing, we would not 
provide much details of the model here. 
In this model, due to the leptons inside the loop, 
the kinetic mixing of $\gamma$ and $\zp$ can be
derived as

\begin{align}
&\varepsilon^{\rm HVL}(q^2)=\Pi(q^{2})
=
\begin{minipage}{4cm}
\unitlength=1cm
\begin{picture}(3.5,1.5)
\put(0,0){\includegraphics[width=3.5cm]{gamm-Z-all}}
\end{picture}
\end{minipage}
\nonumber
\\
%%%%%
&=
\begin{minipage}{4cm}
\unitlength=1cm
\begin{picture}(3.5,1.5)
\put(0,0.05){\includegraphics[width=3.5cm]{gamm-Z-min}}
\end{picture}
\end{minipage}
+\,\,\,\,\,\,\begin{minipage}{4cm}
\unitlength=1cm
\begin{picture}(3.5,1.5)
\put(0,0.05){\includegraphics[width=3.5cm]{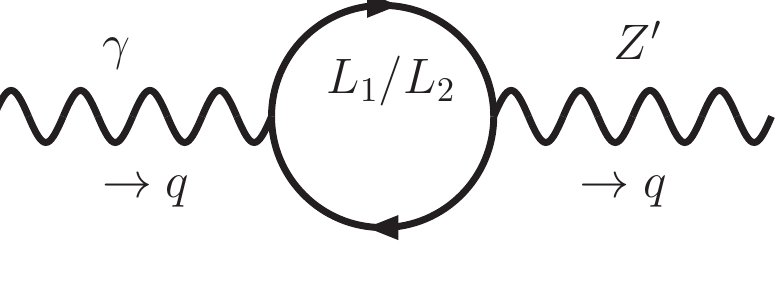}}
\end{picture}
\end{minipage}
\nonumber
\\
%%%%%
&=
\frac{8 e g^\prime}{(4\pi)^2}\int_{0}^{1}\ x(1-x)\Big[\ln\frac{m_\tau^2-x(1-x)q^2}{m_\mu^2-x(1-x)q^2} 
\nonumber
\\
&+\ln\frac{m_{L_2}^2-x(1-x)q^2}{m_{L_1}^2-x(1-x)q^2}\Big]\ dx.
\label{eq:mixing-hl}
\end{align}
Here $m_{L_1},\ m_{L_2}$  are the masses of $L_1$ and $L_2$.
When the momentum transfer $q^2\ll m_{L_1/ L_2}$, which is considered in this work, 
the mixing can be simplified as
\be
\varepsilon^{\rm HVL}(q^2,r)=\varepsilon^{\rm min}(q^2)+\frac{e g_{Z^\prime}}{6\pi^2}
\ln r, 
\label{eq:epstsl}
\ee
where $r={m_{L_2}}/{m_{L_1}}$ is 
the mass ratio of $L_1$ and $L_2$.
%For the sake of convenience, we lable this models as ``Model I" in the following.

\subsection{The $U(1)_{\lmt}$ model with extra heavy charged scalars}
In the $U(1)_{\lmt}$ extension of the SM, we add two extra scalars ($S_1, S_2$) with  
electric charge of $e$ and charged under $U(1)_{\lmt}$ opposite in sign \cite{Banerjee:2018mnw}. 
Similarly, due to charged leptons and extra scalars contributions induced at one-loop level, the $\gamma-\zp$ kinetic mixing can be given as

\begin{align}
&\varepsilon^{\rm HCS}(q^2)=\Pi(q^{2})
=
\begin{minipage}{4cm}
\unitlength=1cm
\begin{picture}(3.5,1.5)
\put(0,0){\includegraphics[width=3.5cm]{gamm-Z-all}}
\end{picture}
\end{minipage}
\nonumber
\\
%%%%%
&=
\begin{minipage}{4cm}
\unitlength=1cm
\begin{picture}(3.5,1.5)
\put(0,0.05){\includegraphics[width=3.5cm]{gamm-Z-min}}
\end{picture}
\end{minipage}
+\,\,\,\,\,\,\
\begin{minipage}{4cm}
\unitlength=1cm
\begin{picture}(3.5,1.5)
\put(0,0.05){\includegraphics[width=3.5cm]{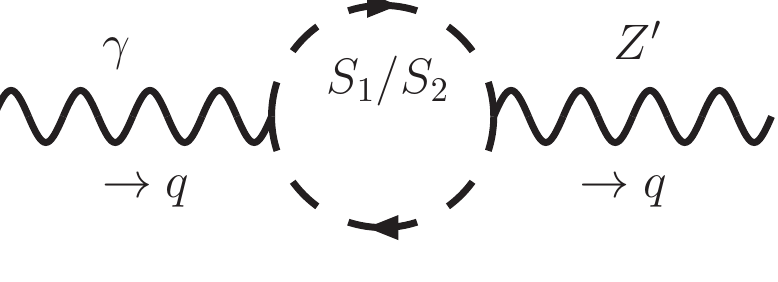}}
\end{picture}
\end{minipage}
\nonumber
\\
%+\,\,\,\,\,\,\
&+
\begin{minipage}{4cm}
\unitlength=1cm
\begin{picture}(3.5,1.5)
\put(0,0){\includegraphics[width=3.5cm]{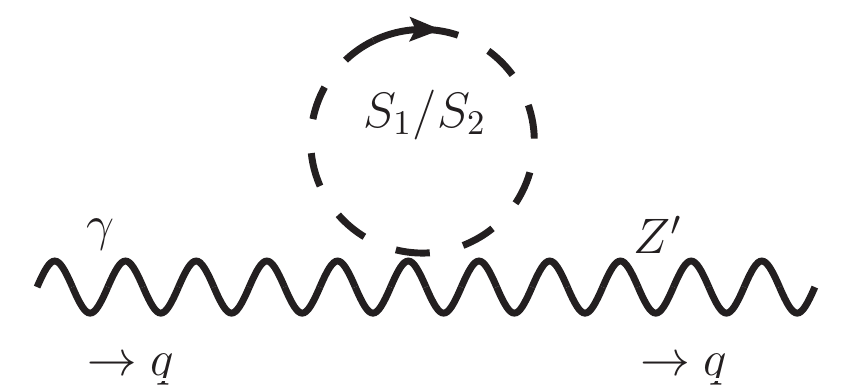}}
\end{picture}
\end{minipage}
\nonumber
\\
%%%%%
&=
\frac{8 e g^\prime}{(4\pi)^2}\int_{0}^{1}dx\ x(1-x)\ln\frac{m_\tau^2-x(1-x)q^2}{m_\mu^2-x(1-x)q^2} 
\nonumber \\
&+ \frac{2 e g^\prime}{(4\pi)^2}\int_{0}^{1}dx\ (1-2x)^2\ln\frac{m_{S_2}^2-x(1-x)q^2}{m_{S_1}^2-x(1-x)q^2}
\label{eq:mixing-hs}
\end{align}
Here $m_{S_1}$ and $m_{S_2}$ are the masses of extra charged scalars ($S_1$ and $S_2$).
We mainly focus on the gauge kinetic mixing, thus much details of the model are not provided here.

In this work, we consider the momentum transfer always $q^2\ll m_{S_1/S_2}$, thus the 
mixing can be also written as 
\be
\varepsilon^{\rm HCS}(q^2,r)=\varepsilon^{\rm min}(q^2)+\frac{e g_{Z^\prime}}{24\pi^2}
\ln r, 
\label{eq:epssusy}
\ee
where $r={m_{S_2}/m_{S_1}}$ is the mass ratio of $S_2$ and $S_1$.

\begin{figure*}[htbp]
\begin{center}
\includegraphics[width=0.45\textwidth]{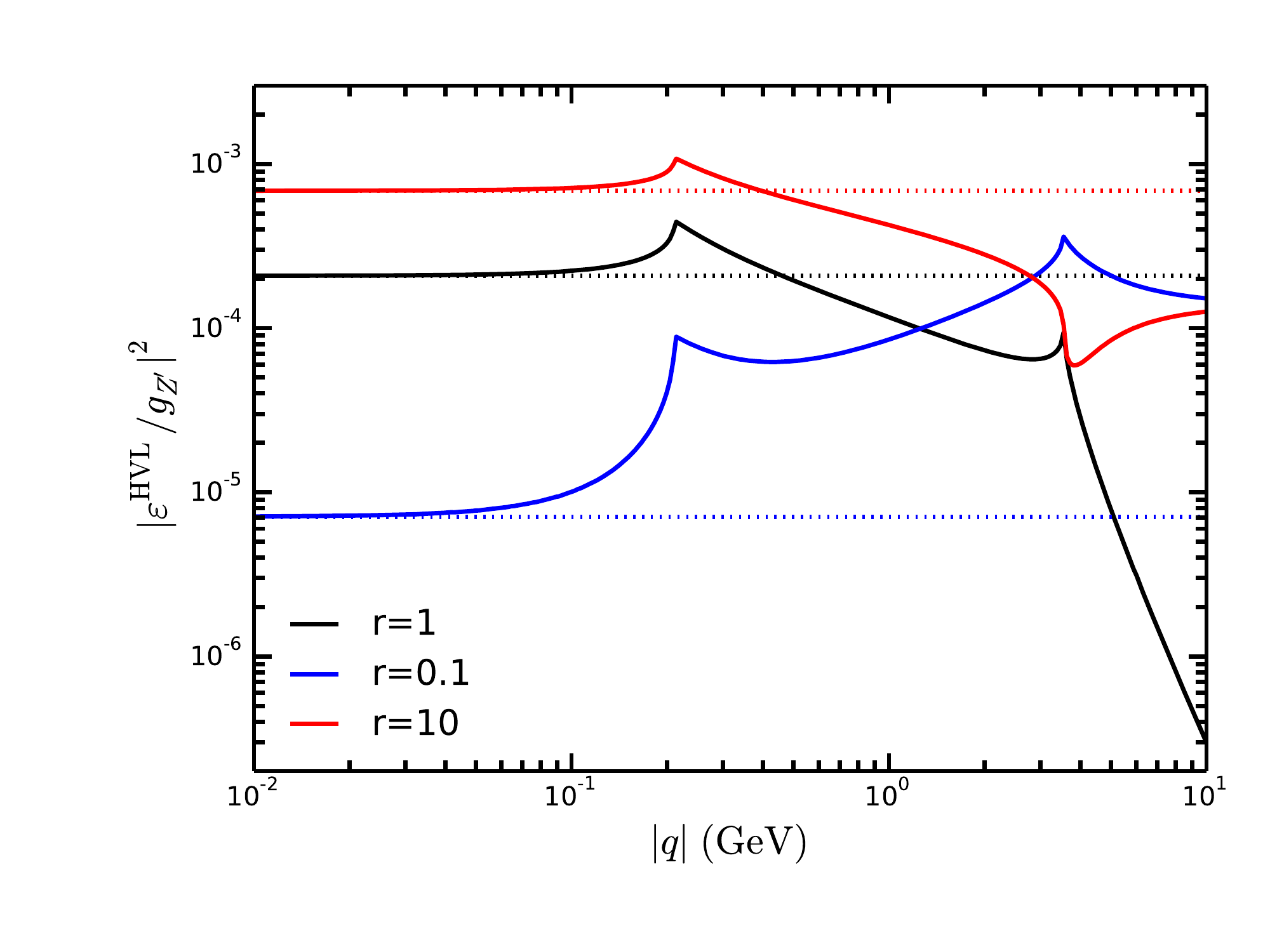}
\includegraphics[width=0.45\textwidth]{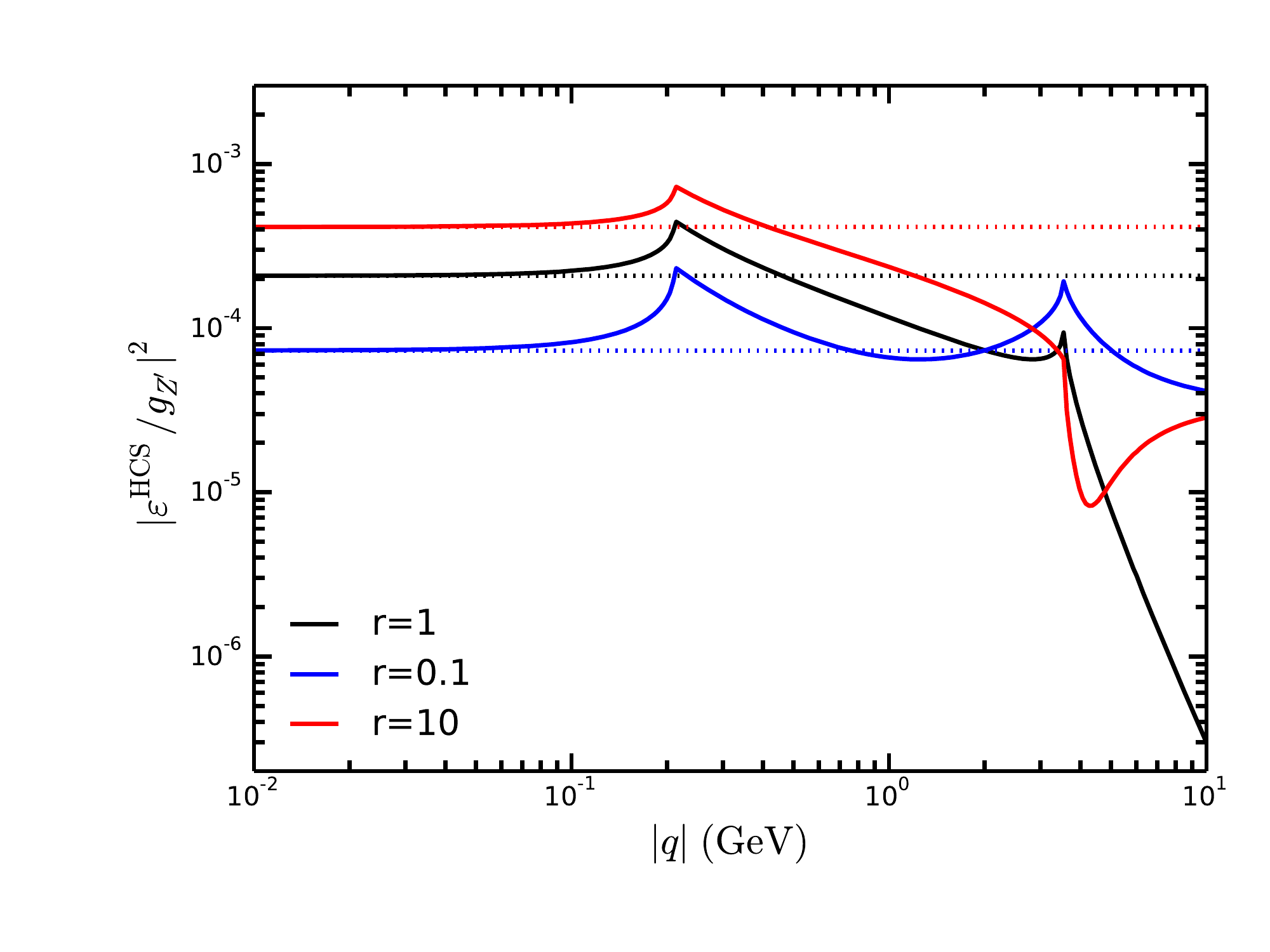}
%\hspace{0.2cm}
%\includegraphics[width=0.18\textwidth]{Figs/FeynDia-eevva}
\caption{The square of the 
kinetic mixing $|\varepsilon/g_{Z^\prime}|^2$ as a 
function of the momentum transfer $|q|$ with the mass ratio $r=0.1, 1$ and 10 for $U(1)_{\lmt}$ model with extra heavy vector-like leptons (Left) or charged scalars (Right). The horizontal dotted lines are the same situations but for the case of $\varepsilon(q^2=0)$, which are shown as a comparison.}
\label{fig:epswithq}
\end{center}
\end{figure*}

In Fig.\ref{fig:epswithq}, we present the square of the 
kinetic mixing $|\varepsilon^{\rm HVL,HCS}/g_{Z^\prime}|^2$ as a 
function of the momentum transfer $|q|$ with $r=0.1, 1$ and 10.
The horizontal dotted lines are the same situations but for the case of $\varepsilon(q^2=0)$, which are shown as a comparison.
When $r=1$, the contribution for the kinetic mixing due to additinal leptons or scalars vanishes, and the results
will become same as those in the minimal $U(1)_{\lmt}$ model, i.e., 
$\varepsilon^{\rm HVL}(q^2,1)=\varepsilon^{\rm HCS}(q^2,1)=\varepsilon^{\rm min}(q^2)$.
In the minimal $U(1)_{\lmt}$ model, $|\varepsilon/g_{Z^\prime}|^2$ has two peaks
at the position of $|q|=m_\mu$ and $|q|=m_\tau$, and drops quickly with 
the increment of $|q|$ when  $|q|> m_\tau$.
This feature distinguishes the phenomenology of the $U(1)_{\lmt}$ model 
from the dark photon models with a constant value of the kinetic mixing.
%%%%% %%%%% %%%%% %%%%% %%%%%
%%%%% %%%%% %%%%% %%%%% %%%%%

\begin{figure*}[htbp]
\begin{center}
\includegraphics[width=0.45\textwidth]{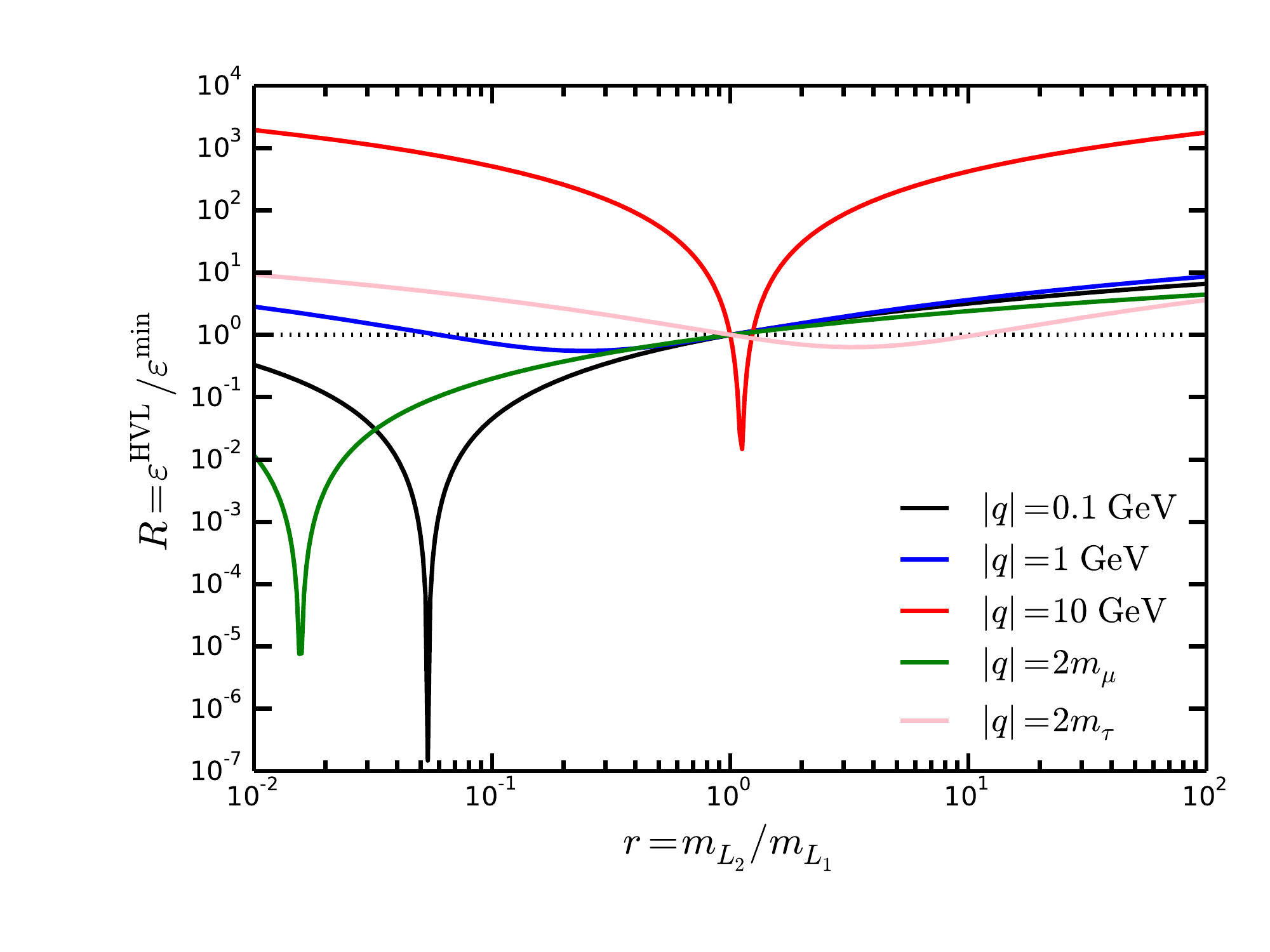}
\includegraphics[width=0.45\textwidth]{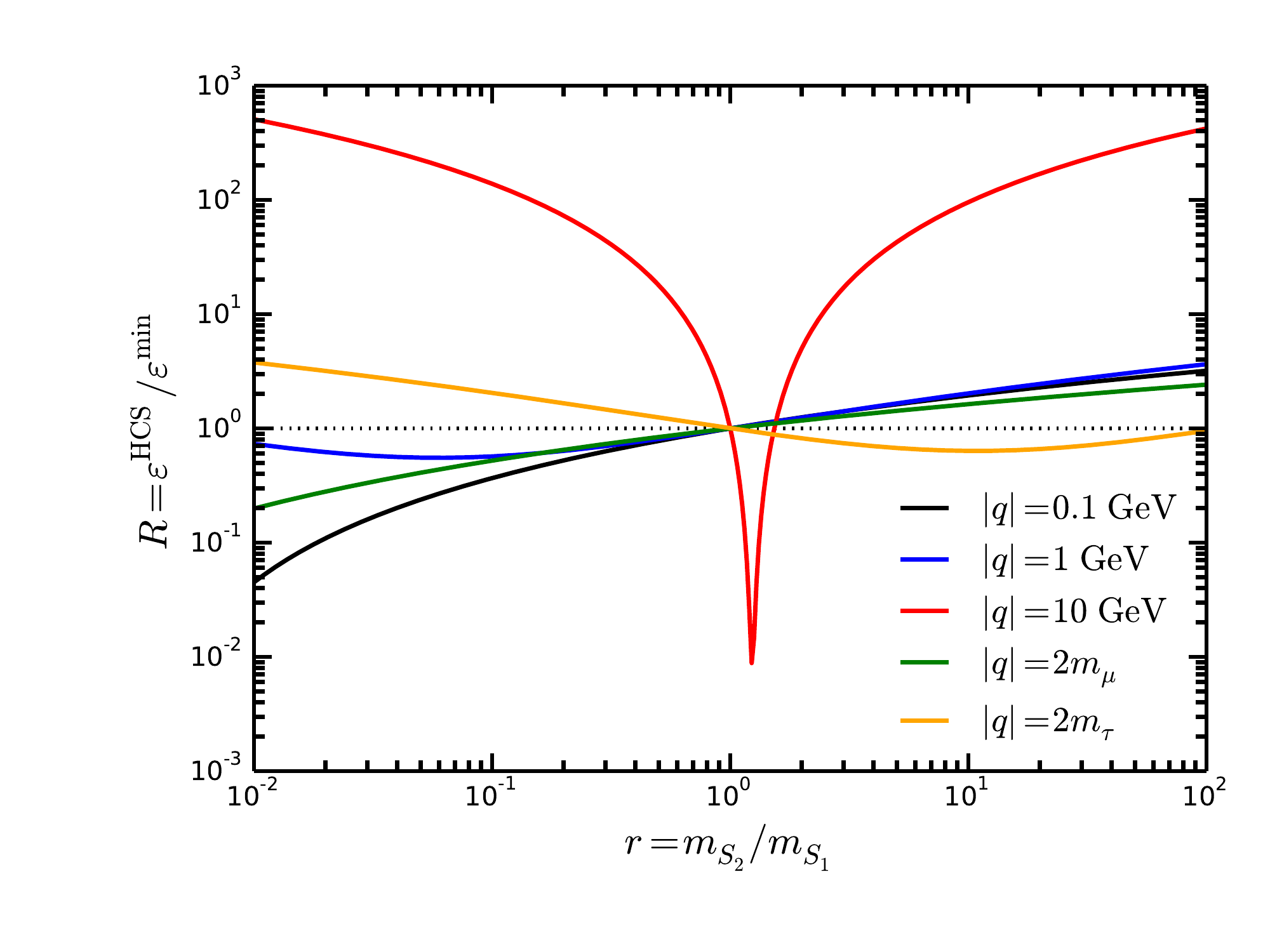}
%\hspace{0.2cm}
%\includegraphics[width=0.18\textwidth]{Figs/FeynDia-eevva}
\caption{The ratio $R=|\varepsilon^{\rm TSL, SUSY}|^2/|\varepsilon^{\rm min}|^2$ of the kinetic mixing between $U(1)_{\lmt}$ model with extra heavy vector-like leptons (Left) or charged scalars (Right) and
the minimal $U(1)_{\lmt}$ model as the funcion of  mass ratio $r$
with momentum transfer $|q|=0.1\GeV,\ 1\GeV,\ 10\GeV,\ 2m_\mu,\ 2m_\tau$.}
\label{fig:epsratiowithq}
\end{center}
\end{figure*}

We also present the dependence of the kinetic mixing ratio 
$R=|\varepsilon^{\rm HVL/HCS}|^2/|\varepsilon^{\rm min}|^2$ 
between the  $U(1)_{\lmt}$ model 
with two singlet vectorlike leptons or with two charged scalars
and the minimal $U(1)_{\lmt}$ model on the mass ratio $r$  in Fig.\ref{fig:epsratiowithq}.
There we consider five typical momentum transfers $|q|=0.1\GeV,\ 1\GeV,\ 10\GeV,\ 2m_\mu,\ 2m_\tau$.
It can be seen that,  the additional lepton or scalar contributions could be significant, and
the results are distinctly different from those of the minimal $U(1)_{\lmt}$ model.
Though the additional leptons and scalars cannot be detected directly due to their heavy mass, they can provide significant contributions to the kinetic  mixing.

%\paragraph{Invisible decay of $\zp$.}
\subsection{Decay modes of $\zp$}
Since the $\zp$ direct couples with the leptons of second and third generation, it can decay into a pair of
neutrinos, and also may decay into muon and tau leptons if kinematic allowed.
In addition, since $\zp$ provides possible scenarios of  dark matter, there can be the channel $\zp\to\chi\bar{\chi}$.
The decay widths of $\zp$ are given by,
\begin{align}
\Gamma(Z' \rightarrow \nu_\ell\bar{\nu_\ell}) 
& = \frac{g_{Z'}^{2}}{24\pi}m_{\zp},
\\
%%%%%
\Gamma(Z' \rightarrow \ell^{+}\ell^{-})
& =  \frac{g_{Z'}^{2}}{12\pi}m_{\zp}
\left[
1+\frac{2m_{\ell}^{2}}{m_{\zp}^{2}}
\right]
\sqrt{1-\frac{4m_{\ell}^{2}}{m_{\zp}^{2}}},
\\
\Gamma(Z' \rightarrow \chi\bar{\chi})
& =  \frac{g_{D}^{2}}{12\pi}m_{\zp}
\left[
1+\frac{2m_{\chi}^{2}}{m_{\zp}^{2}}
\right]
\sqrt{1-\frac{4m_{\chi}^{2}}{m_{\zp}^{2}}},
\end{align}
where $\ell = \{\mu, \tau\}$, $g_D$ is the coupling constant of the $\zp$ with dark matter, and $g_D\gg g_{Z'}$ is assumed.
We ignore the channel $\zp\to e^+e^-$ since it is suppressed by the kinetic mixing.
Since neutrinos and dark matter are invisible at particle detectors, we take the $\zp$ invisible decay as
$\Gamma(Z' \rightarrow {\rm INV})=\Gamma(Z' \rightarrow \nu\bar{\nu})+\Gamma(Z' \rightarrow \chi\bar{\chi})$, whose decay ratio can be expressed as
\begin{small}
\begin{align}
{\rm Br}(Z'\!\! \rightarrow\!\! {\rm INV})
&\!=\!
\begin{cases}
\!1, & \hspace{-0.2cm}(m_{\zp} < 2m_\mu\, {\rm or}\ m_{\zp}> 2m_\chi),
\vspace{0.2cm}
\\
\!\cfrac{\Gamma(Z' \rightarrow \nu\bar{\nu})}
{\displaystyle
\sum_{f=\nu, \mu}
\hspace{-0.1cm}
\Gamma(Z' \rightarrow f\bar{f})},
& \hspace{-0.2cm}(2m_\mu\! < m_{\zp}\! < 2m_\tau\ \! {\rm and}\ m_{\zp}\!< \!2m_\chi),
\vspace{0.2cm}
\\
\!\cfrac{\Gamma(Z' \rightarrow \nu\bar{\nu})}
{\displaystyle
\sum_{f=\nu,\mu,\tau}
\hspace{-0.25cm}
\Gamma(Z' \rightarrow f\bar{f})},
& \hspace{-0.2cm}(2m_\tau < m_{\zp}\, {\rm and}\ m_{\zp}< 2m_\chi).
\end{cases}
\end{align}
\end{small}

%%%%% %%%%% %%%%% %%%%% %%%%%
%%%%% %%%%% %%%%% %%%%% %%%%%

\section{Existing constraints}

\begin{figure}[htbp]
	\begin{center}
		\includegraphics[scale=0.40]{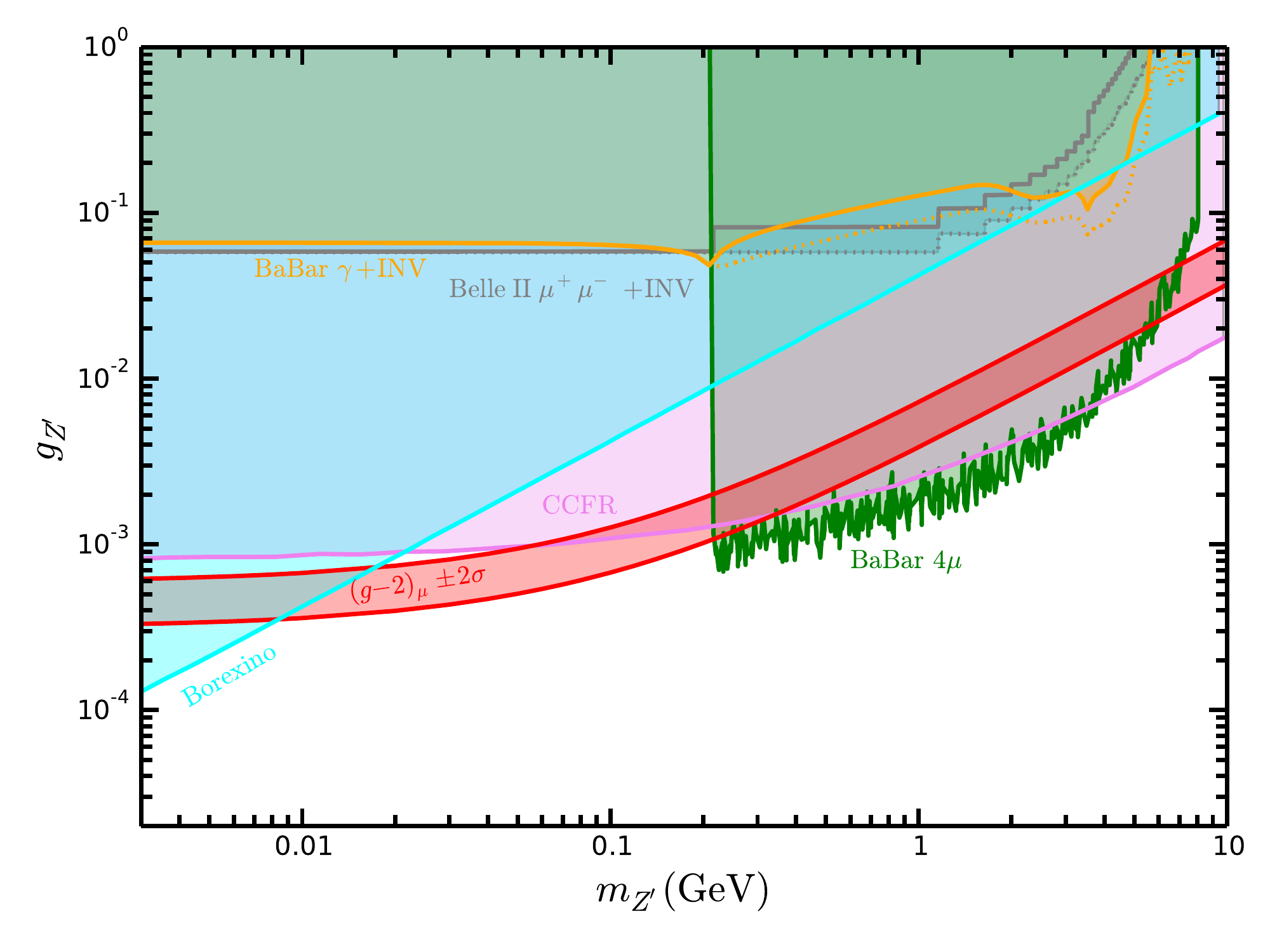}
		\caption{
Summary for the $\mzp-g_{\zp}$ parameter space of the mininal $U(1)_{\lmt}$ model,
{where $Z^\prime$ has no additional decay channel to dark sector}.
The shaded regions show the exisiting bounds excluded by CCFR experiment
in neutrino trident production \cite{Altmannshofer:2014pba}, by Borexino detector in neutrino-electron scattering \cite{Abdullah:2018ykz},
by BaBar in the reactions $\epem\to\mu^+\mu^-\zp, \zp\to\mu^+\mu^-$ with 514 fb$^{-1}$ data \cite{TheBABAR:2016rlg}
and  $\epem\to\gamma \zp, \zp\to {\rm INV}$ with 53 fb$^{-1}$ data \cite{Lees:2017lec}, 
and by Belle II in the process $\epem\to\mu^+\mu^-\zp, \zp\to {\rm INV}$ with 276 pb$^{-1}$ data \cite{Adachi:2019otg}.
%\zx{We assume ${\rm BR}(\zp\to{\rm INV})\simeq1$ for the invisible decay of $\zp$.}
{The dotted lines indicate ${\rm BR}(\zp\to{\rm INV})\simeq1$ cases.}
The red band indicate the allowed region at $2\sigma$ from the experimental measurements of muon magnetic momentum.}
		\label{fig:curren-limits}
	\end{center}
\end{figure}

In this section, we summarize the existing constraints relevant to the parameter regions we are
interested for the minimal $U(1)_{\lmt}$ model from various experiments as follows:
\begin{itemize}
\item{\it Muon anomalous magnetic moment.}
The significant discrepancy between the experimental measurement and the SM prediction in the 
magnetic moment of the muon remains one of the
largest anomalies in particle physics \cite{Tanabashi:2018oca}:
\bea\label{eq:g-2}
\Delta a_{\mu}^{Z^{\prime}}&\equiv& a_{\mu}^{\mathrm{exp}}-a_{\mu}^{\mathrm{SM}} \nonumber \\
&=&\left(261 \pm 61_{\mathrm{exp}} \pm 48_{\mathrm{the}}\right) \times 10^{-11},
\eea
where the errors are from experiment and theory prediction, respectively.
We require the contribution in Eq.(\ref{eq:g-2}) to be within $2\sigma$
that leads to
\begin{equation}
103 \lesssim \Delta a_{\mu}^{Z^{\prime}} \times 10^{11} \lesssim 420.
\end{equation}
The minimal $U(1)_{\lmt}$ model, was first introduced to address the discrepancy,
which can provide a new interaction with muons.
An extra contribution to $a_{\mu}$ arises solely from a one-loop diagram involving $\zp$,
which can be giving by 
\begin{equation}
a_{\mu}^{\zp}=\frac{g_{{\zp}}^{2}}{8 \pi^{2}} \int_{0}^{1} \frac{2 m_{\mu}^{2} x^{2}(1-x)}{x^{2} m_{\mu}^{2}+(1-x) m_{\zp}^{2}} d x.	
\end{equation}
The parameter region on which the $\zp$ contribution in the minimal $\lmt$ model 
resolves the discrepancy in the muon anomalous magnetic moment at
$2\sigma$ is indicated with the red band in Fig. \ref{fig:curren-limits}.
%%%%%%%%%%%%%%%%%%%%%%%%%%%%%%%%%%%%%%%%%%%%%%%%
\item {\it Neutrino trident production.} 
The neutrino trident production is a muon neutrino scattering  off the Coulomb field of a target 
nucleus ($N$), producing two muons in the final state,
$\nu N \rightarrow \nu N \mu^{+}\mu^{-}$. Besides the SM $Z$ boson, in the $U(1)_{\lmt}$ model, the 
$\zp$ boson can also contribute to this process, which 
can offer a sensitive search for the light $\zp$ boson \cite{Altmannshofer:2014pba,Magill:2016hgc}.
The measurements for the cross section have been reported by CCFR,
which obtain the result  $\sigma_{\mathrm{CCFR}} / \sigma_{\mathrm{SM}}=0.82 \pm 0.28$. 
The bound is depicted in Fig. \ref{fig:curren-limits} and taken from Ref. \cite{Altmannshofer:2014pba}.
%%%%%%%%%%%%%%%%%%%%%%%%%%%%%%%%%%%%%%%%
\item {\it Neutrino-electron scattering.}  
The neutrino-electron elastic scattering processes can probe $U(1)_{\lmt}$ gauge boson,
since $U(1)_{\lmt}$ gauge boson can contribute through the kinetic mixing. 
The most stringent constraints 
come from the  Borexino solar neutrino experiment.
Limits for $U(1)_{\lmt}$ gauge boson have been derived 
in Refs.\cite{Araki:2017wyg, Bauer:2018onh} 
by converting existing bounds 
on $U(1)_{B-L}$ models \cite{Harnik:2012ni} using earlier 
Borexino $^7$Be data \cite{Bellini:2011rx}. 
The bounds are updated in Ref.\cite{Abdullah:2018ykz} 
using the recently-released Borexino measurement of $^7$Be neutrinos\cite{Agostini:2017ixy}. 
We show them in Fig. \ref{fig:curren-limits}. 	
%%%%%%%%%%%%%%%%%%%%%%%%%%%%%%%%%%%%%%%%%%%%%%%%%%%%%
\item {\it $\zp$ production associated with muon pair.
%	 $\zp$ production without the kinetic mixing at electron-positron colliders.
}
Via the direct coupling to $\mu$,
$\zp$ can be produced  at $\epem$ colliders in 
the process $\epem\to\mu^+\mu^-\zp$. Babar experiment 
has reported the bounds using 514 fb$^{-1}$ data collected
in the reaction $\epem\to\mu^+\mu^-\zp, \zp\to\mu^+\mu^-$
for $m_{\zp} > 2 m_\mu$ \cite{TheBABAR:2016rlg}. 
Recently, Belle II experiment perform the first searches 
for the invisble decay of a $\zp$ in the process $\epem\to\mu^+\mu^-\zp, \zp\to {\rm INV}$ 
using 276 pb$^{-1}$ collected \cite{Adachi:2019otg}, which can touch the region of 
$m_{\zp} < 2 m_\mu$. 

%	\item {\it Belle II  $\epem\to\mu^+\mu^-\zp, \zp\to {\rm invisible}$ channel search.}
%%%%%%%%%%%%%%%%%%%%%%%%%%%%%%%%%%%%%%%%%%%%%%%%%%%%%%%%%%%%
\item {\it $\zp$ production associated with SM photon.
}
At $\epem$ colliders, the $\zp$ boson can also be produced associated 
with SM photon via the kinetic mixing  in the process $\epem\to \gamma \zp$ \cite{Lees:2017lec}. 
The search for invisible decays of dark photon has been preformed at 
BaBar experiment using the single-photon events with 53 fb$^{-1}$ data.
We translate the constraints for dark photon to
$U(1)_{\lmt}$ gauge boson $\zp$ using 
\be\label{eq:dp2zp}
\varepsilon_{\rm DP}^2\to |\varepsilon|^2{\rm Br}(\zp\to {\rm INV}),
\ee
where $\varepsilon_{\rm DP}$ is the photon and dark photon kinetic mixing parameter in the dark photon model,
and $\varepsilon$ is the $\gamma-Z^\prime$ kinetic mixing in the $U(1)_{\lmt}$ model.
\end{itemize}

{In Fig. \ref{fig:curren-limits}, we asume $\zp$ does not decay into dark sector,
i.e., $\Gamma(\zp \rightarrow {\rm INV})=\Gamma(\zp \rightarrow \nu\bar{\nu})$.  
The  ${\rm BR}(\zp\to{\rm INV})\simeq 1$ cases are also shown as dotted line for a visual display.}
Taking the constraints above into account, a narrow window of the $\mzp-g_{\zp}$ parameter
region in the minimal $U(1)_{\lmt}$ model desired by the muon anomalous magnetic moment,
\begin{equation}
10\,\mathrm{MeV} \lesssim M_{Z^{\prime}} \lesssim 210\, \mathrm{MeV}, \quad 4 \times 10^{-4} \lesssim g_{X} \lesssim 10^{-3},
\end{equation}
is still allowed.

\section{Searching for $U(1)_{\lmt}$ gauge boson at electron colliders }
	
\begin{figure}[htbp]
	\begin{centering}
		\includegraphics[width=0.5\columnwidth]{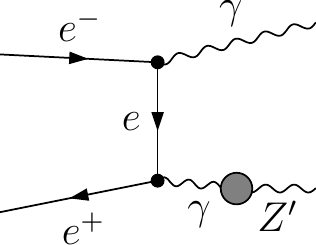}
		\caption{The Feynman diagrams for the production of an on-shell $Z^\prime$ and a photon.
%			, in which we assume the
%			$A^\prime$ subsequently decays to lighter dark matter.
		}
		\label{fig:eeaA} 
	\end{centering}
\end{figure}

At the electron colliders, the production of $Z^\prime$ can be associated with a SM photon  
through the kinetic mixing in the process $e^+e^-\to \gamma \zp$,  whose diagrams are shown in Fig.\ref{fig:eeaA}.
Subsequently, the produced $\zp$ boson can decay 
into charged leptons, a pair of neutrinos or light dark matter. 
In this paper, we focus on the $\zp$ invisible decay channel
$\zp\to{\rm INV}$, including $\zp\to\nu\bar\nu$ and $\zp\to\chi\bar{\chi}$,
to probe $\zp$ boson via the monophoton searches $\epem\to\gamma \zp \to \gamma +{\rm INV}$ at electron colliders. 
We assume that the decay width of the $\zp$ is negligible compared to the experimental resolution, 
which justifies the use of the narrow width approximation.

%%%%%%%%%%%%%%%%%%%%%%%%%%%%%%%%%%%%%%%%%%%%%%%%%%%%%%%%%%%%%%%
%%%%%%%%%%%%%%%%%%%%%%%%%%%%%%%%%%%%%%%%%%%%%%%%%%%%%%%%%%%%%%%

In the monophoton signature at electron colliders, the major backgrounds (BGs) from SM contain two types:
irreducible and reducible BG. 
The irreducible monophoton BG comes from the process \eevva,
where $\nu$ is the three neutrinos.
The reducible monophoton BG arises 
from the electromagnetic processes $e^+e^-\to \gamma +\slashed{X}$,
where $\slashed{X}$ denotes other visible particles but undetected due to the limitations of 
the detector acceptance.  
We discuss the reducible BG in detail later 
for each experiment, since it strongly depends on the 
angular coverage of the detectors.

The differential cross section for an on-shell $Z^\prime$ and a photon production 
process $e^+e^-\to \gamma Z^\prime$  is 
\cite{Essig:2009nc}
\be
\frac{d\sigma_{\gamma\zp}}{dz_\gamma} = 
\frac{2\pi \alpha^2|\varepsilon(m_{\zp}^{2})|^{2}}{s}\left(1-\frac{m_{\zp}^2}{s}\right)
\frac{1+z_\gamma^2+\frac{4sm_{\zp}^2}{(s-m_{\zp}^2)^2}}{(1+z_\gamma)(1-z_\gamma)},
\ee
where $\alpha$ is the fine structure constant, $z_\gamma\equiv\cos\theta_\gamma$ with $\theta_\gamma$ 
being the relative angle between the electron beam axis and the photon momentum in the center-of-mass (CM) frame,
$s$ is the square of the CM energy, $m_{Z^\prime}$ is the mass of the $U(1)_{\lmt}$ gauge boson.
The photon energy $E_\gamma$ in the CM frame is related to the $Z^\prime$ mass as 
\be
E_\gamma = \frac{s-m_{\zp}^2}{2\sqrt{s}}.
\label{eq:egma}
\ee
The cross section after integrating the polar angle $\theta_\gamma$ 
is given as \cite{Essig:2009nc}
\bea
\label{eq:xs-signal}
\sigma_{\gamma \zp}&=&\frac{2\pi \alpha^2 |\varepsilon(m_{\zp}^{2})|^{2}}{s}\left(1-\frac{m_{\zp}^2}{s}\right) \nn \\
&\times&\left[\left(1+\frac{2sm_{\zp}^2}{(s-m_{\zp}^2)^2}\right){\cal Z}
-z_\gamma^{\rm max}+z_\gamma^{\rm min}\right],
\eea
where
\be
{\cal Z}=\ln\frac{(1+z_\gamma^{\rm max})(1-z_\gamma^{\rm min})}{(1-z_\gamma^{\rm max})(1+z_\gamma^{\rm min})} .
\ee

\section{Belle \uppercase\expandafter{\romannumeral2}}\label{sec:belle2}

At Belle II, photons and electrons can be detected in the 
Electromagnetic Calorimeter (ECL), 
which is made up of three segments:
forward endcap with $12.4^{\circ} < \theta < 31.4^{\circ}$,
barrel  with $32.2^{\circ} < \theta < 128.7^{\circ}$, 
and backward endcap $130.7^{\circ} < \theta < 155.1^{\circ}$ 
in the lab frame \cite{Kou:2018nap}.
At Belle II, the reducible BG for monophoton singnature
consists of two major parts:
one is mainly due to the lack of polar angle  coverage 
of the ECL near the beam directions, which is referred to 
as the ``bBG''; 
the other one is {mainly} due to the gaps between 
the three segments in the ECL detector,  
which is referred to as the ``gBG''.

The bBG  comes from the electromagnetic processes $e^+e^-\to \gamma +\slashed{X}$,
manily including $e^+e^-\to\slashed{\gamma}\slashed{\gamma}\gamma$ and \eesesea,
where all the other final state particles  except the detected photon
are emitted along the beam directions with 
$\theta>155.1^{\circ}$ or $\theta<12.4^{\circ}$ in the lab frame. 
At Belle II, we adopt the detector cuts for the final detected photon 
(hereafter the ``{\it pre-selection cuts}"):
$12.4^{\circ}<\theta_\gamma<155.1^{\circ}$ in the lab frame.

\begin{figure*}[htbp]
\begin{center}
\includegraphics[width=0.45\textwidth]{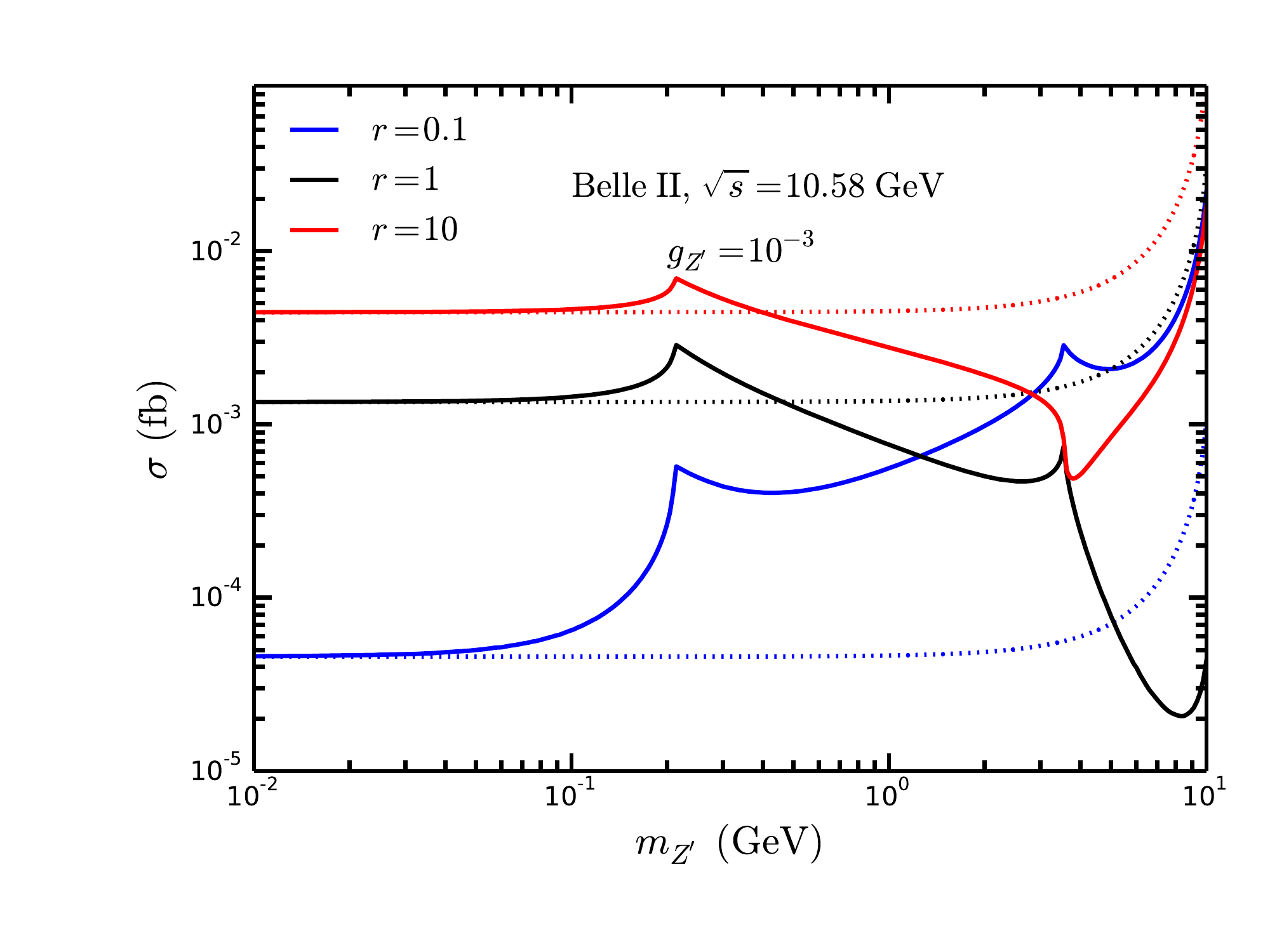}
\includegraphics[width=0.45\textwidth]{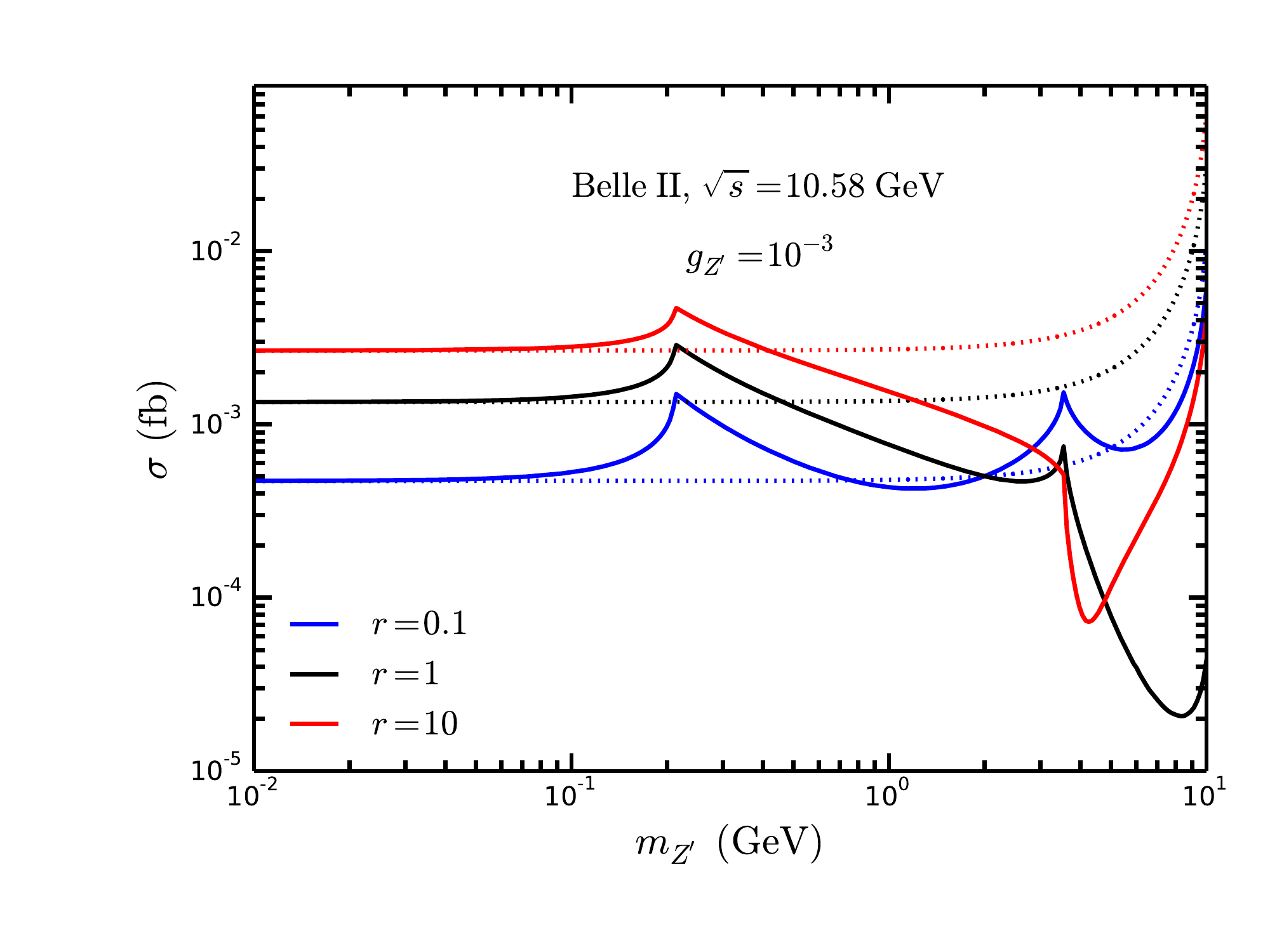}
\caption{
The cross sections of the process $e^+e^-\to \gamma Z^\prime$  at Belle II with $\sqrt s = 10.58$ GeV
after the ``{\it pre-selection cuts}" for $\lmt$ model 
with extra heavy vector-like leptons (Left) or charged scalars (Right).
The horizontal dotted lines are the same situations but for the case of $\varepsilon(q^2=0)$, which are shown as a comparison.
}
\label{fig:sigma-belle2}
\end{center}
\end{figure*}

In Fig.\ref{fig:sigma-belle2}, we show the production rates of the process $e^+e^-\to \gamma Z^\prime$ 
in $U(1)_{\lmt}$ models after the ``{\it pre-selection cuts}" for the photon at Belle II with $\sqrt s = 10.58$ GeV.
The  dotted lines correspond to the case of constant $\varepsilon(q^2=0)$, which are shown as a comparison.
We can see that, with constant $\varepsilon(q^2=0)$, the cross sections all increase with 
the increment of the mass of $\zp$. In the minimal $U(1)_{\lmt}$ model, the production rates for the process $e^+e^-\to \gamma \zp$ at Belle II generally drop but exist two peaks at the positions of $m_\mu$ and $m_\tau$ when
$\mzp<8.5$ GeV, while raise at the tail of the plotted region.

%%%%% %%%%% %%%%% %%%%% %%%%%
%%%%% %%%%% %%%%% %%%%% %%%%%

For the Belle II detector, which is asymmetric, 
the maximum energy of the monophoton events 
in the bBG in the CM frame,
$E_\gamma^m$,  is given by \cite{Liang:2019zkb}
(if not exceeding $\sqrt{s}/2$) 
\be
E_\gamma^m(\theta_\gamma) = 
\frac{ \sqrt{s}(A\cos\theta_1-\sin\theta_1)}
{A(\cos\theta_1-\cos\theta_\gamma)-(\sin\theta_\gamma+\sin\theta_1)},
\label{eq:bBG}
\ee 
where  
all angles are given in the CM frame, 
and $A=(\sin\theta_1-\sin\theta_2)/(\cos\theta_1-\cos\theta_2)$, 
with $\theta_1$ and $\theta_2$ being 
the polar angles corresponding to 
the edges of the ECL detector.
In order to remove the above bBG, we use the detector 
cut $E_\gamma > E_\gamma^m$ (hereafter the {\it``bBG cuts"}) for the final monophoton .

\begin{figure*}[htbp]
	\begin{center}
		\includegraphics[width=0.45\textwidth]{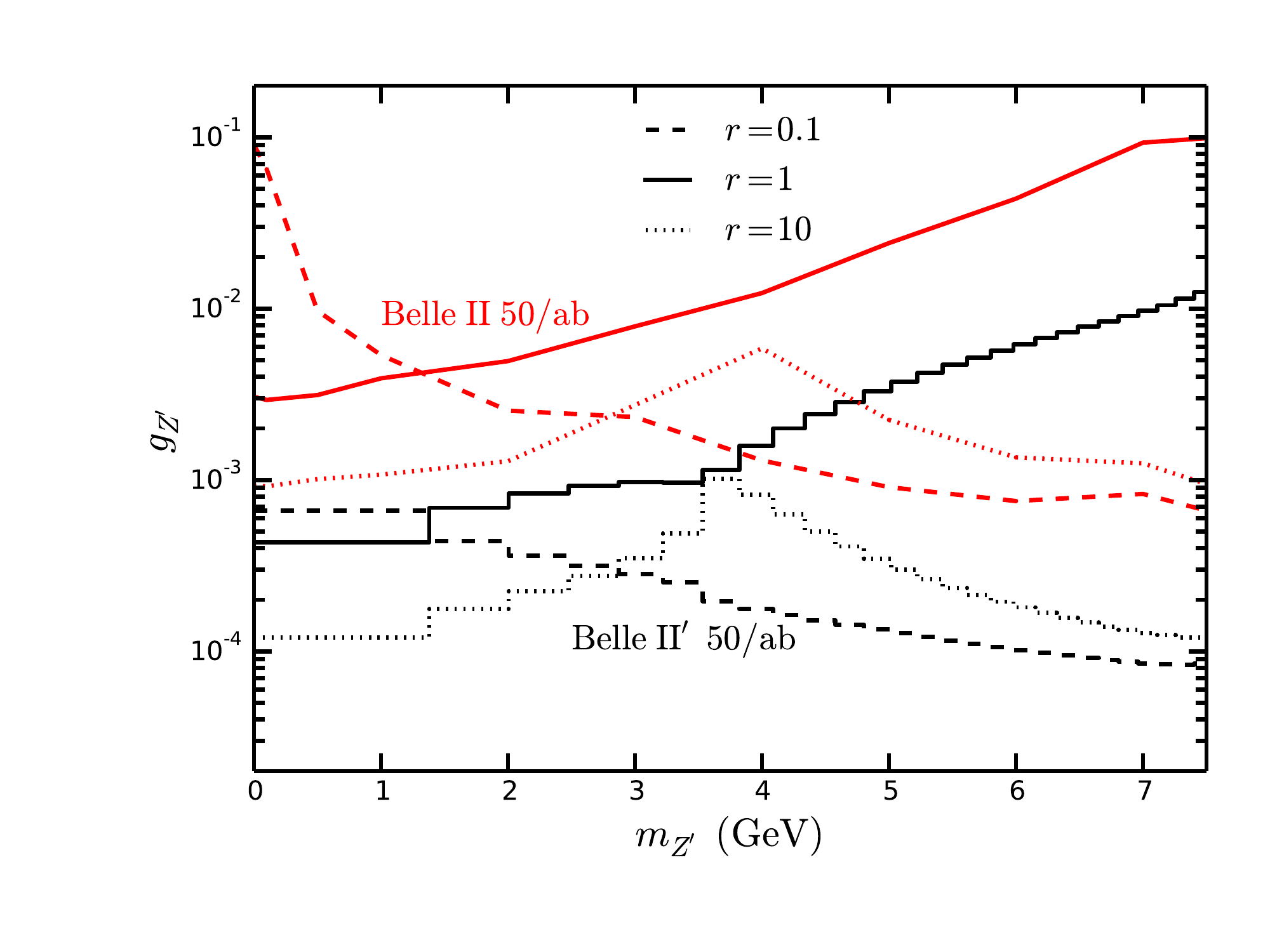}
		\includegraphics[width=0.45\textwidth]{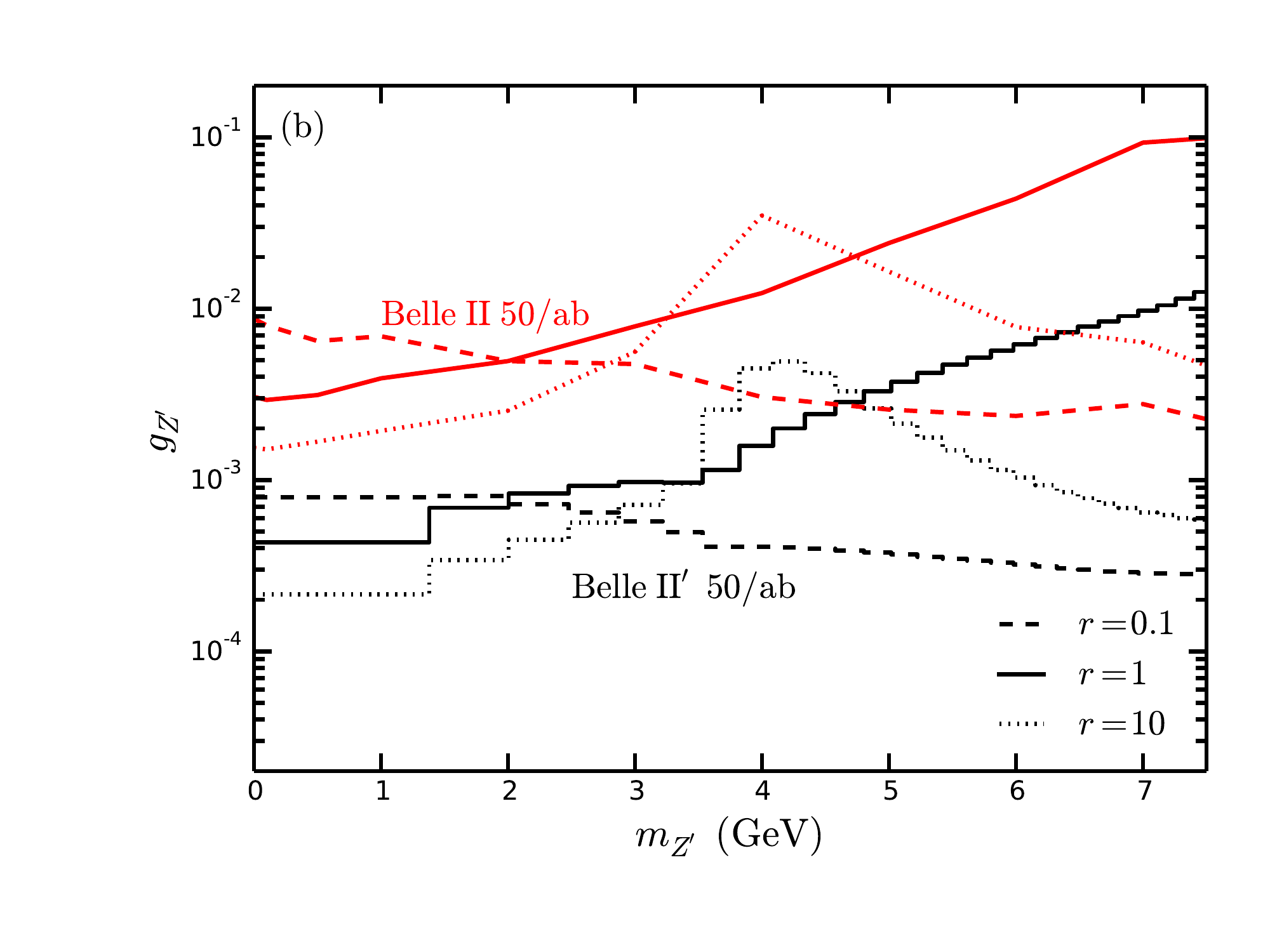}
		\caption{
			Sensitivity limit for $g_{\zp}$ at Belle-II experiment with 50 ab$^{-1}$ to search for dark photon decaying into 
			light dark matter based on the simulation in Ref.\ \cite{Kou:2018nap}, lablled as ``Belle II", red color.
			The expected 95\% C.L. exclusion limits on $g_{\zp}$ via monophoton searches at 50 ab$^{-1}$ Belle-II with gBG omitted	after {\it``optimized cut"}, labeled as Belle-II$^\prime$, black color.
			For $\lmt$ model with extra heavy vector-like leptons (Left) or charged scalars (Right)
			in the cases of $r=0.1$ (dashed), 1 (solid) and 10 (dotted).}
		\label{fig:eps-belle2}
	\end{center}
\end{figure*}

The gBG for the monophoton singnature have been simulated in the 
Ref.\ \cite{Kou:2018nap} to search for dark photons decaying into light dark matter.
The projetced upper limits on the kinetic mixing of dark photon and SM photon $\varepsilon_{\rm}$
for a 20 fb$^{-1}$ Belle II dataset are present there.
The constranint for the kinetic mixing between $U(1)_{\lmt}$ gauge boson and SM photon $\varepsilon$ 
can be translated from the dark photon using
Eq.\ (\ref{eq:dp2zp}).
We scale the expected sensitivity ${\cal S}(g_{\zp})$ to the planned full of integrated luminosity 
of 50 ab$^{-1}$ at Belle II using ${\cal S}(g_{\zp}) \propto \sqrt[4]{\cal L}$.
Then the corresponding constraint based on the simulation in 
Ref.\ \cite{Kou:2018nap} from 20 fb$^{-1}$ to 50 ab$^{-1}$ 
can be simply projected by a factor of $\sqrt[4]{50/ {\rm ab}\over 20/ {\rm fb} }$, 
%$({50/ {\rm ab}\over 20/ {\rm fb} })^{1/4}$
which is present in  Fig.\ref{fig:eps-belle2} and the invisible decay 
ratio ${\rm Br(\zp\to INV)}\simeq 1$ is assumed.
It is shown that 
the sensitivity for $g_{\zp}$ at Belle-II experiment with 50 ab$^{-1}$ 
via monophoton searches is expected to be worse in the minimal $U(1)_{\lmt}$ model with the increament of $\mzp$,
while become better with extra heavy vector-like leptons (charged scalars) in the case of $r=0.1$ when $\mzp<$7 GeV. 
With $r=10$ in the $U(1)_{\lmt}$ model with extra heavy leptons (scalars),  expected $g_{\zp}$ sensitivity gets improved
when $\mzp\gtrsim4$ GeV and then gets worse.

We further carry out an analysis without gBG taking into
account, to compare with other experiments in which 
detailed simulations with gBG are not available.
We use the {\it``bBG cuts"} to remove the reducible BG
events; this momentum the BG events survived the {\it``bBG cuts"} come from irreducible BG
without gBG considered.
Since the energy of the final photon in the signal process is related to $m_{Z^\prime}$,
in addiction to the {\it``bBG cuts"}, we select final photon in the energy window of 
$| E_\gamma -(s-m_{Z^\prime}^2)/(2\sqrt{s})|< \sigma_E/2$ (hereafter the {\it``optimized cut"}) to enhance the discovery sensitivity, 
where $\sigma_E$ is detector energy resolution for the photon.
At Belle II,  $\sigma_E / E = 4\% (1.6\%)$ at 0.1 (8) GeV \cite{Kou:2018nap}
and we take $\sigma _{E}=128$ MeV conservatively.
In Fig. \ref{fig:eps-belle2}, we present the expected 95\% confidence level (C.L.)  exclusion 
limits on $g_{\zp}$ by considering the irreducible BG only
after {\it``optimized cut"}, which is labeled as Belle-II$^\prime$.
We define 
$\chi^{2}(\varepsilon) \equiv S^2/(S+B)%
$ \cite{Yin:2009mc}, 
where $S$ ($B$) is the number of events in the signal (BG) processes. 
The 95\% C.L. upper bound  
on $g_{\zp}$ is obtained by solving 
$\chi^{2} (\epsilon_{95})-\chi^{2} (0)=2.71$,
and assuming photon detection efficiency as 95\% \cite{Kou:2018nap}.
One can see that if we don't consider the ``gBG" and apply the {\it``optimized cut"},
the Belle II experiment with 50 ab$^{-1}$ via monophoton searches is expected to be sensitive to
the parameter region with $m_{\zp}\lesssim 1.2$  GeV and $g_{\zp}\gtrsim 4\times 10^{-4}$
in the minimal $\lmt$ model, which can be improved by almost 1 order of magnitude comparing 
with considering the ``gBG".

%%%%% %%%%% %%%%% %%%%% %%%%%
%%%%% %%%%% %%%%% %%%%% %%%%%	

\section{BESIII and STCF}
	
\begin{figure*}[htbp]
\begin{center}
\includegraphics[width=0.45\textwidth]{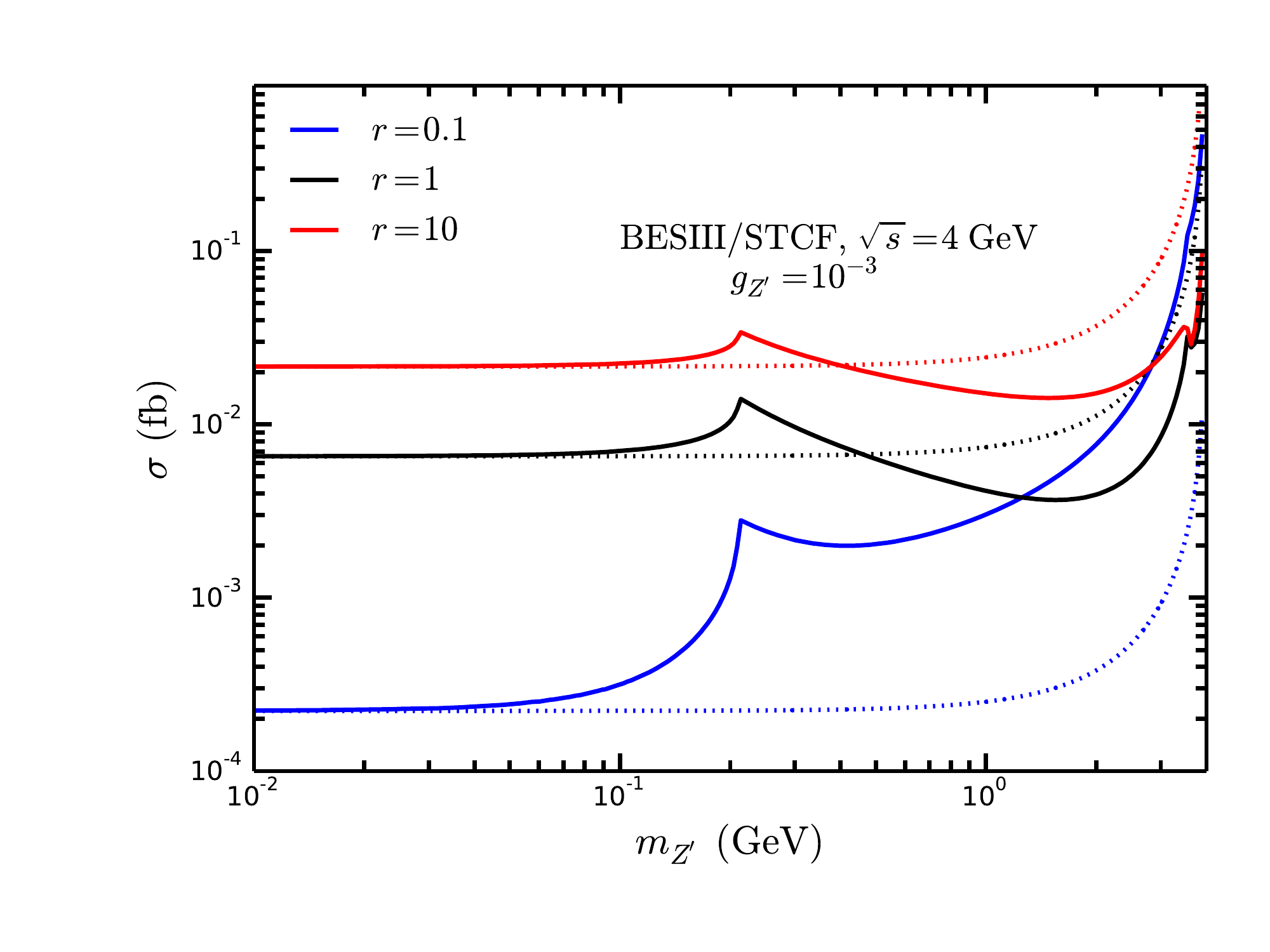}
\includegraphics[width=0.45\textwidth]{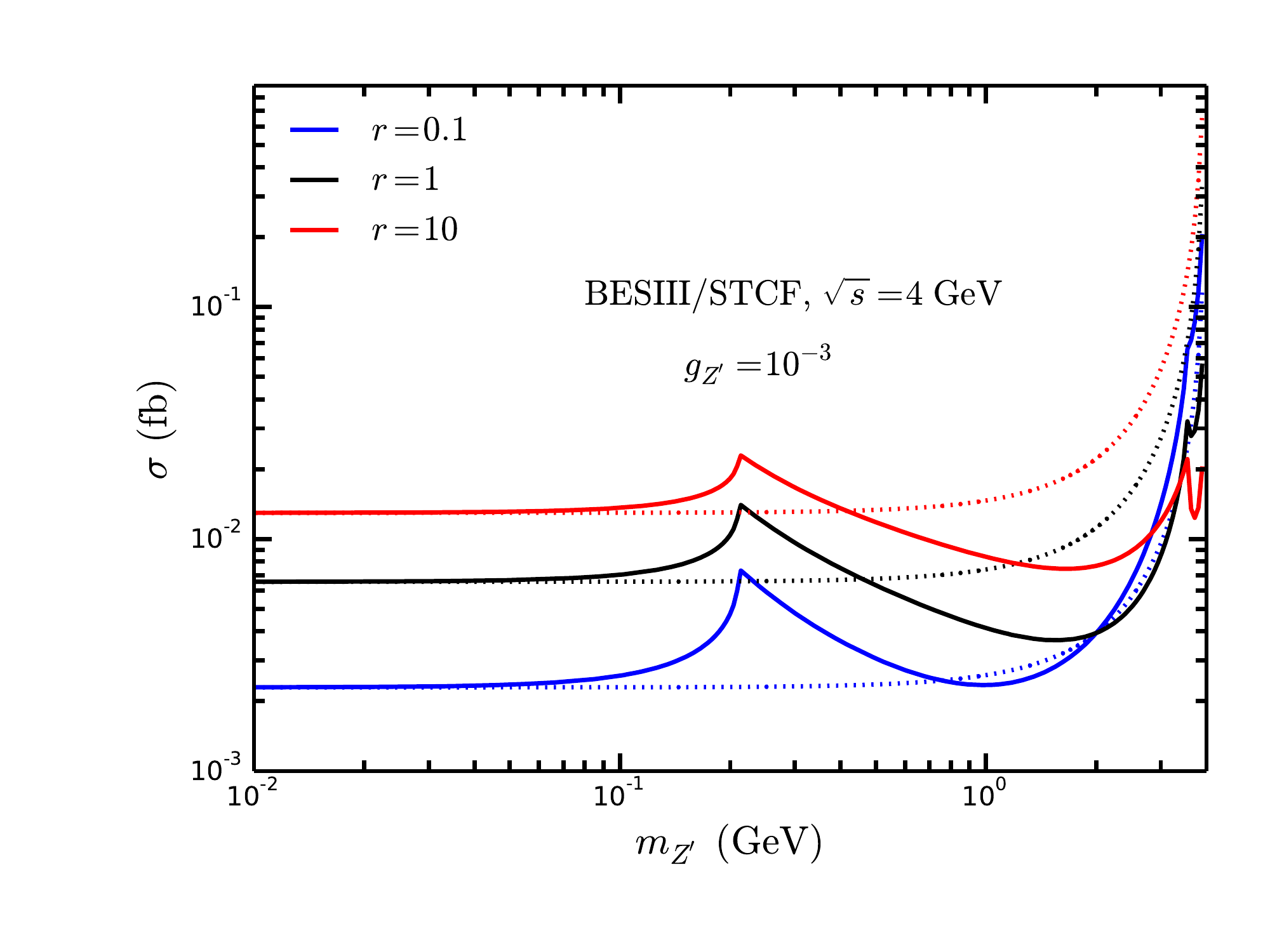}
\caption{The cross sections of the process $e^+e^-\to \gamma Z^\prime$  at BESIII or STCF with $\sqrt s = 10.58$ GeV
after the ``{\it pre-selection cuts}" for $\lmt$ model 
with extra heavy vector-like leptons (Left) or charged scalars (Right).
The horizontal dotted lines are the same situations but for the case of $\varepsilon(q^2=0)$, which are shown as a comparison.
}
\label{fig:xs-besiii}
\end{center}
\end{figure*}

At BESIII and STCF, for the final state photons, we adopt the {\it ``preselection cuts"} by 
BESIII Collaboration
\cite{Ablikim:2017ixv}: 
$E_\gamma>$ 25 MeV with $| \cos \theta |<0.8$ 
or $E_\gamma>$ 50 MeV with $0.86<| \cos \theta |<0.92$.
In Fig. \ref{fig:xs-besiii}, we present the cross section of the the process $\epem\to\gamma\zp$ at BESIII and STCF 
with $\sqrt s = 4$ GeV in $U(1)_{\lmt}$ models after the ``{\it pre-selection cuts}".
The  dotted lines correspond to the case of constant $\varepsilon(q^2=0)$, which are shown as a comparison.
One can see that, the cross section always increases for larger $\mzp$ in $U(1)_{\lmt}$ models with extra 
heavy leptons or scalars in the case of $r=0.1$, while there is
a twist near $\mzp=2m_\tau$ in the case of  $r=1$ and $r=10$.

%%%%% %%%%% %%%%% %%%%% %%%%%
%%%%% %%%%% %%%%% %%%%% %%%%% 

At BESIII and STCF, which are symmetric, 
the maximum energy of the monophoton events in the
bBG in the CM frame, 
$E_\gamma^m$,  is given by \cite{Liu:2019ogn}
\be
E_\gamma^m(\theta_\gamma) = 
\sqrt{s}\left(\frac{\sin\theta_\gamma}{\sin\theta_b}\right)^{-1},
\label{eq:bBG-besiii}
\ee 
where $\cos\theta_b$ is the polar angle corresponding to the edge
of the detector.
Taking into account the coverage of 
MDC, EMC, and TOF, we have $\cos\theta_b=0.95$ at the BESIII \cite{Liu:2018jdi}. 
We further demand $E_\gamma > E_\gamma^m$ for the final monophoton to remove the reducible 
BG (hereafter the {\it``bBG cuts"}).

At BESIII, the photon energy resolution of the EMC 
$\sigma_E / E=2.3 \% / \sqrt{E / \mathrm{GeV}} \oplus 1 \%$ \cite{Asner:2008nq}, 
and we take $\sigma _{E}=40$ MeV for all energy conservatively.
At the BESIII, photon reconstruction efficiencies are
all more than 99\% \cite{Ablikim:2011kv}, we assume them to be 100\% in our paper.
For the EMC at STCF, we assume the same energy resolution and reconstrunction efficiencies with BESIII to present a preliminary projection limit,
because of the similarity of the two experiments.
We take $\sigma _{E}=25\ (40,\ 50)$ MeV for $\sqrt s=2,\ (4,\ 7)$ GeV.
In addition to the {\it``bBG cuts"}, we select final photon in the energy window of 
$| E_\gamma -(s-m_{Z^\prime}^2)/(2\sqrt{s})|< \sigma_E/2$ (hereafter the {\it``optimized cut"}) to enhance the discovery sensitivity.

\begin{figure*}[htbp]
\begin{center}
\includegraphics[width=0.45\textwidth]{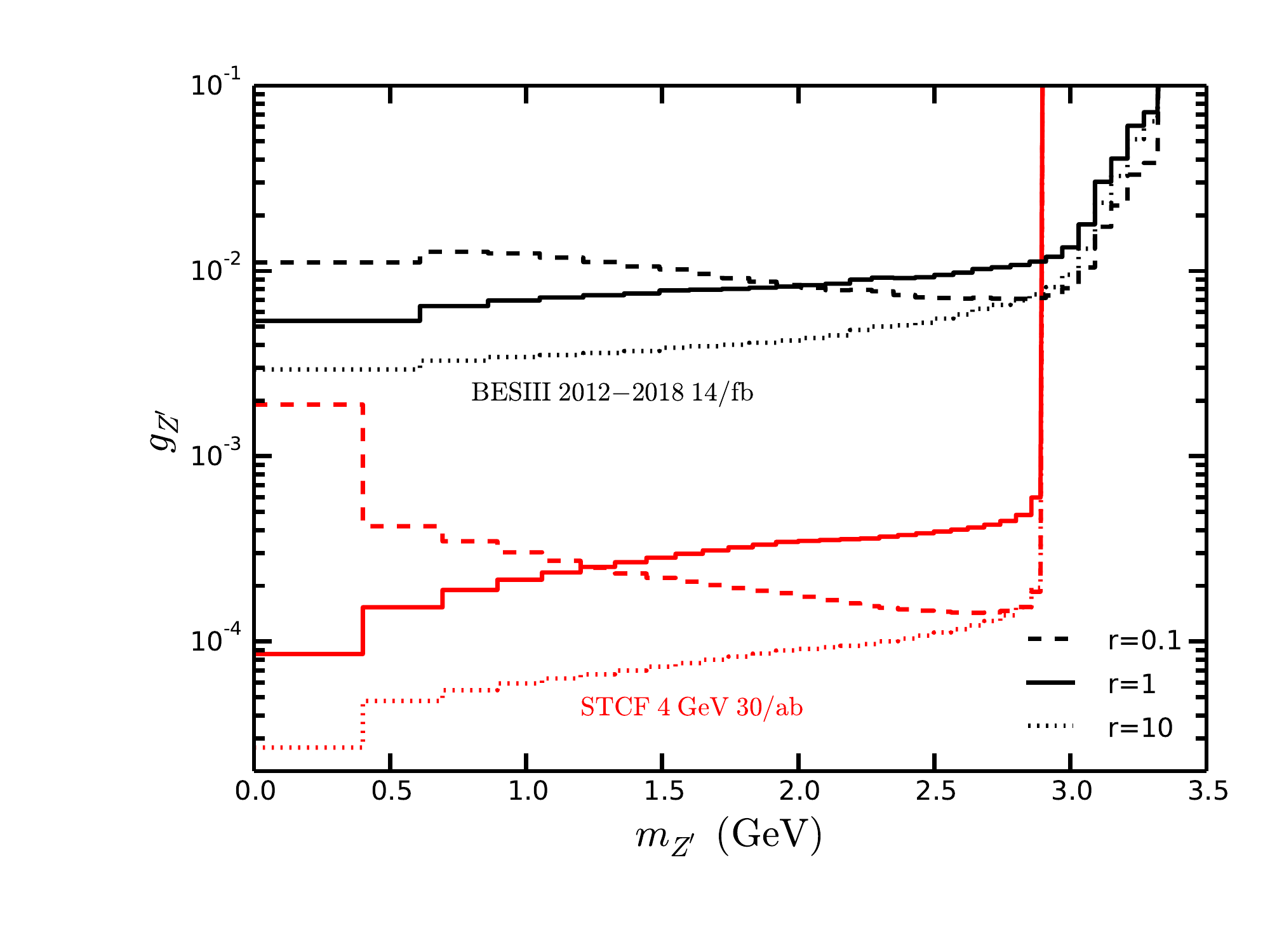}
\includegraphics[width=0.45\textwidth]{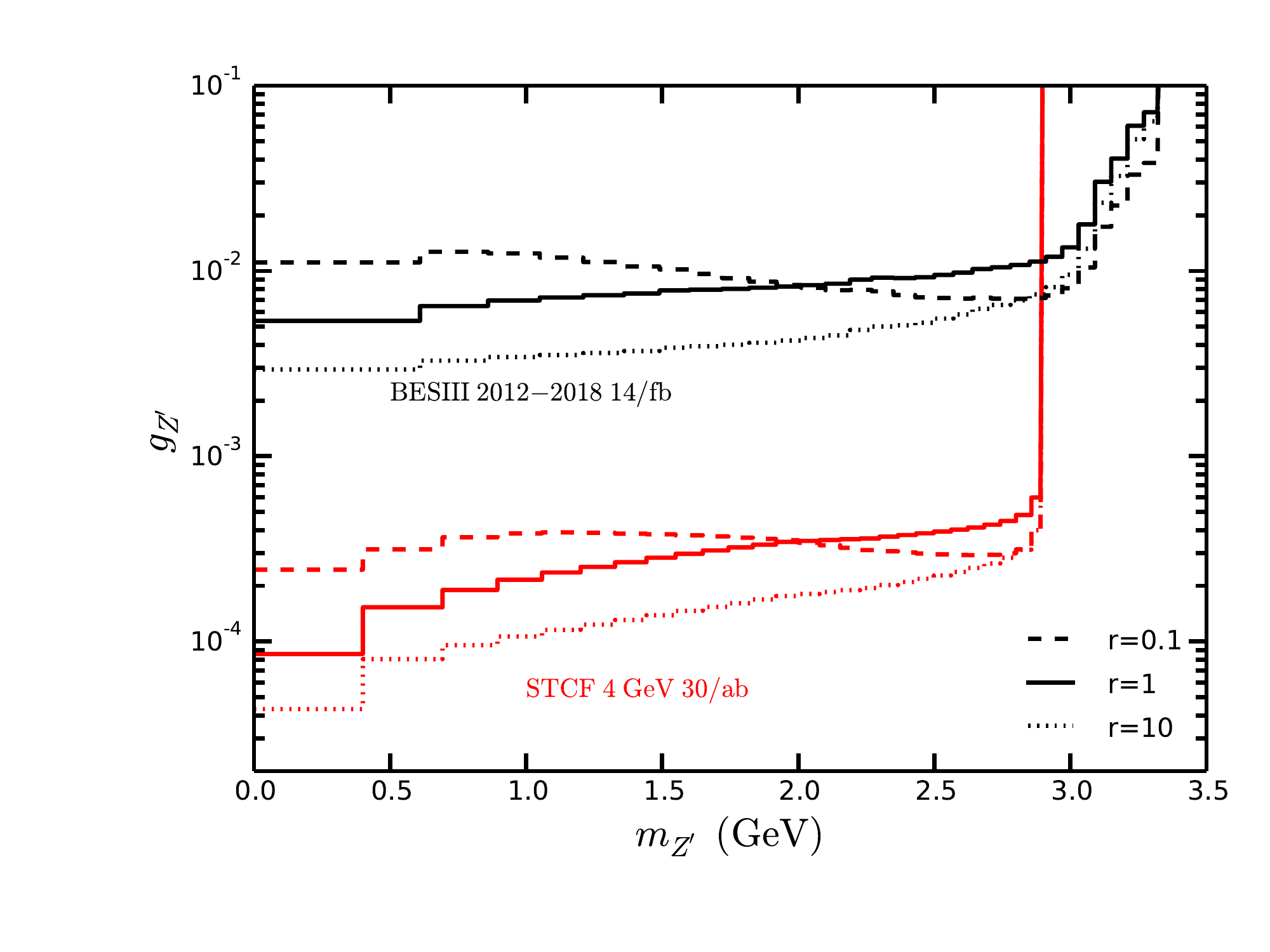}
\caption{The expected 95\% C.L. exclusion limits on $g_{\zp}$ via monophoton searches after {\it``optimized cut"} 
at BESIII with 14 fb$^{-1}$ (black color) and 4 GeV STVF with 30 fb$^{-1}$ (red color).
For $\lmt$ model with extra heavy vector-like leptons (Left) or charged scalars (Right)
in the cases of $r=0.1$ (dashed), 1 (solid) and 10 (dotted).
}
\label{fig:eps-stcf}
\end{center}
\end{figure*}

At BESIII, since 2012 monophoton trigger has been implemented and the corresponding data luminosity 
reach about 14 fb$^{-1}$ with the CM energy from 2.125 GeV to 4.6 GeV \cite{Zhang:2019wnz}. 
We define 
$\chi^{2}_{\rm tot} (\varepsilon)=\sum_{i} \chi_{i}^{2} (\varepsilon) ,$
where 
$\chi^{2}_i (\varepsilon) \equiv S_i^2/(S_i+B_i) $
for each BESIII colliding energy. 
The 95\% C.L.\ upper bound on $g_{\zp}$ from BESIII is 
obtained by demanding $\chi_{\rm tot}^2(\varepsilon_{95})-\chi_{\rm tot}^2(0)=2.71$.
In Fig.\ \ref{fig:eps-stcf}, we present 
the corresponding results for the $U(1)_{\lmt}$ models with extra vector-like 
leptons and charged scalars 
in cases of $r=0.1,\ 1,\ 10$
via monophoton searches
at  BESIII with 14 fb$^{-1}$ and at 4 GeV STCF with 30 ab$^{-1}$, respectively.
The invisible decay ratio of $\zp$ is assumed to be 1.
The constraints on $g_{\zp}$ get  looser with the increament of
$\mzp$ for both two considered models in cases of $r=1,\ 10$ 
at BESIII and 4 GeV STCF, while tighter in cases of $r=0.1$ 
for the $U(1)_{\lmt}$ models with extra leptons (scalars) when 
$\mzp\lesssim 2.7$ GeV ($1.0 \GeV \lesssim \mzp\lesssim 2.7$ GeV) at
4 GeV STCF.

%%%%% %%%%% %%%%% %%%%% %%%%%
%%%%% %%%%% %%%%% %%%%% %%%%%

\section{Results}

\begin{figure*}[htbp]
	\begin{center}
		\includegraphics[scale=0.6]{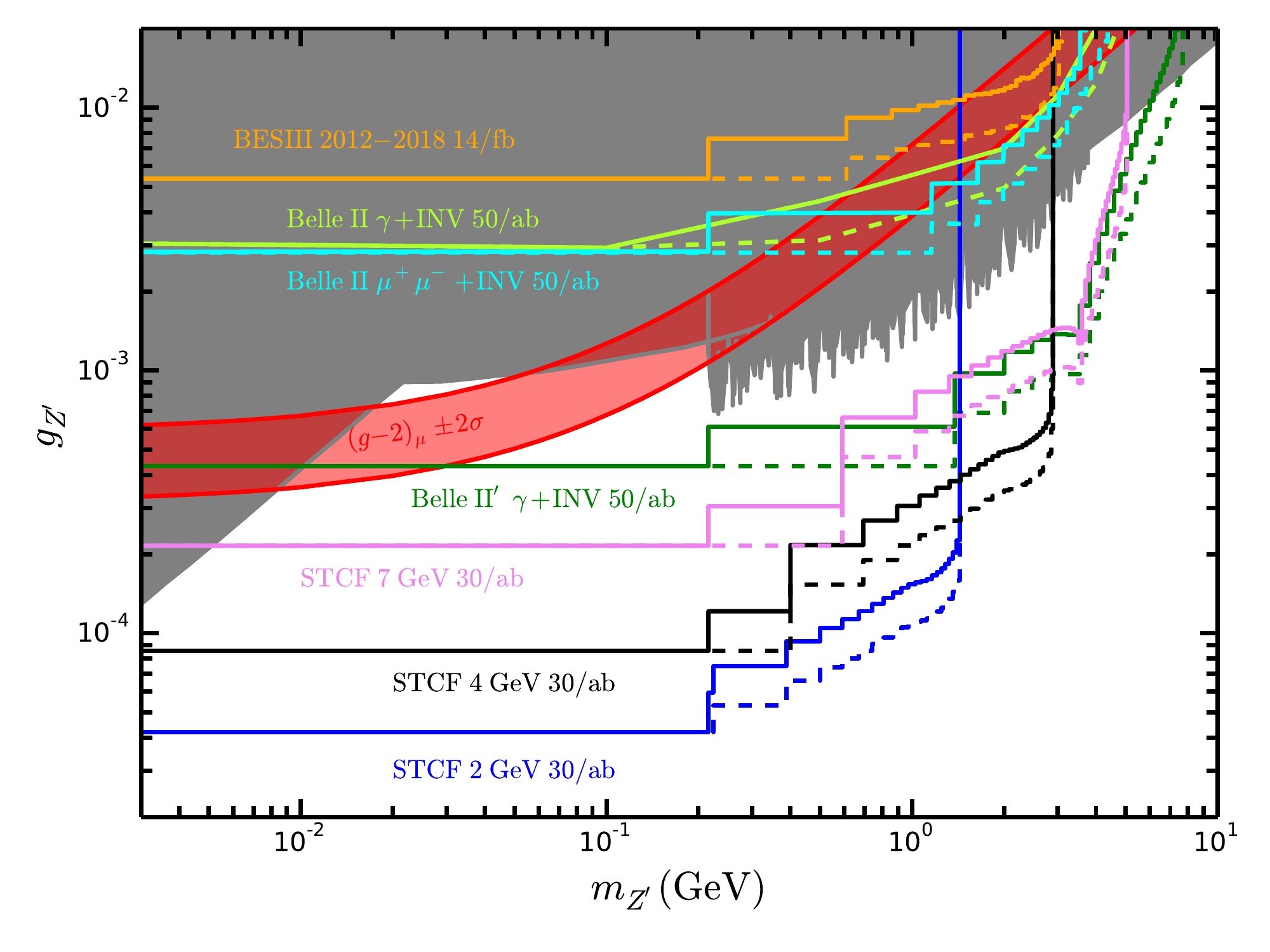}
		\caption{
The sensitivity on
gauge coupling $g_{\zp}$  at Belle II, BESIII and STCF. Notice that we do not 
include the gBG analysis for BESIII and STCF limits.
The solid lines indicate the case of that $\zp$ has no additional decay channel
to dark matter, and the dotted lines indicate  ${\rm Br}(Z' \rightarrow {\rm invisible})\simeq 1$ cases.
The exisiting constraints from different experiments  are presented in the shaded region, and summarized  in Fig.\ref{fig:curren-limits}.
The red band shows the region that could explain the muon anomalous 
magnetic moment $(g-2)_{\mu} \pm 2 \sigma$.
The BESIII limit is obtained with the 14 fb$^{-1}$ luminosity of  monophoton 
trigger collected during 2012-2018. 
The STCF limits are obtained for $\sqrt{s}=2,\ 4,\ 7{\rm\, GeV}$ with the future integrated luminosity of $30\ {\rm ab^{-1}}$. 
The Belle II limits are obtained with future integrated luminosity of $50\ {\rm ab^{-1}}$
with three experiments (See text in detail).
		}
		\label{Fig-Combine-2}
	\end{center}
\end{figure*}

Fig.\ \ref{Fig-Combine-2} summarizes the sensitivity on
gauge coupling $g_{\zp}$  in the minimal $\lmt$ model from  electron 
colliders, including Belle II, BESII, and STCF. 
The solid lines indicate the case of that $\zp$ cannot decay into dark  matter,
i.e., ${\rm Br}(Z' \rightarrow {\rm INV})={\rm Br}(Z' \rightarrow \nu\bar{\nu})$,
and the dotted lines indicate  ${\rm Br}(Z' \rightarrow {\rm INV})\simeq 1$ case.
The exisiting constraints are also presented in the shaded region, and the summary for these 
limits from different experiments can be found in Fig.\ref{fig:curren-limits}.
The red band shows the region that could explain the muon anomalous 
magnetic moment $(g-2)_{\mu} \pm 2 \sigma$.
We present three expected limits with different experiments at Belle II,
\begin{enumerate}
	\item{\it $\gamma+{\rm INV}$ channel with bBG and gBG considered. }
	We translate the constraints on the dark photon from the search of invisible decay at Belle II
	assuming a 20 fb$^{-1}$ dataset \cite{Kou:2018nap},
	where the bBG and gBG are all considered,
	 to $\lmt$ gauge boson using the relation of Eq.(\ref {eq:dp2zp}). Then we scale the constraints to 50 ab$^{-1}$ by a factor of 
	$\left(50\ {\rm ab}^{-1}\over 20\ {\rm fb}^{-1}\right)^{1/4}$
	. This case is labeled as 
	``Belle II $\gamma + {\rm INV}$" in Fig.\ (\ref{Fig-Combine-2}).
	\item{\it $\gamma+{\rm INV}$ channel with only bBG considered. }
	We compute the limits without gBG taking into account as mentioned above.
	The {\it ``bBG cuts"} are applied to remove the reducible BG
	events and only the irreducible BG contribute to the BG events if gBG is not considered. After the {\it ``optimized cut"}, we show the 95\% C.L. upper bound on 
	$g_{\zp}$ at Belle II with the integrated luminosity of 50 ab$^{-1}$ in Fig.\ref{Fig-Combine-2}, which is lablled as ``Belle II$^\prime\ \gamma + {\rm INV}$"
	\item{\it $\mu^+\mu^- +{\rm INV}$ channel. }
	In order to project the sensivity on the $U(1)_{\lmt}$ gauge boson $\zp$ 
	with  $\epem\to\mu^+\mu^-\zp, \zp\to {\rm INV}$ channel in  50 ab$^{-1}$ Belle II experiment,
	we simply scale the recent 276 pb$^{-1}$ results by a factor of $\left(50\ {\rm ab}^{-1}\over 276\ {\rm pb}^{-1}\right)^{1/4}$ for the kinetic mixing, which is lablled as ``Belle II $\mu^+\mu^- + {\rm INV}$"
\end{enumerate}
One observes that on the searches for the invisible decay of $\zp$, the sensitivity 
at 50 ab$^{-1}$ Belle II with $\mu^+\mu^- +{\rm INV}$ channel is slightly better 
with the $\gamma+{\rm INV}$ channel. 
It can also be found that these two results are already excluded by current constraints.
While without the gBG considered in the $\gamma+{\rm INV}$ channel,
the sensitivity can be improved almost 1 order and the gauge coupling constant $g_{\zp}$ down
to about $4.2\times 10^{-4}$ when $\mzp<2m_\mu$,
which still left a thin slice of mass region $\sim (0.01-0.03)$ GeV to explain the moun ($g-2$) anomaly.
The one order of magnitude difference in sensitivity 
between the two Belle II limits via the monophoton search, 
shows that the control on gGB is very important in 
probing the $\zp$ parameter space.

The STCF and BESIII limits are obtained when the BG
due to the gaps in the detectors are neglected,
since BESIII did not  released any analysis about gBG. 
We emphasize that more rigorous BESIII and STCF sensitivities could be obtained
with such gBG anlysis available in the future.
With about 14 fb$^{-1}$ integrated luminosity collected during 2012-2018 \cite{Zhang:2019wnz}
the upper limits from BESIII are exclued by CCFR experiment.
The STCF limits are presented at $\sqrt{s}=2, (4, 7)$ GeV with the integrated luminosity of 30 ab$^{-1}$. 
The future monophoton searches at the STCF experiment operated at $\sqrt{s}=2-7$ GeV can eliminate the moun $g-2$ favored window  when $m_{\zp}\lesssim5$ GeV.
In the low mass region, 2 GeV  STCF provide best sensitivity since 
the signal to BG  ratio increases 
when the colliding energy decreases,
and $g_{\zp}$ can be down
to about $4.2\times 10^{-5}$ when $\mzp<2m_\mu$,
which is improved about 1 order than the monophoton searches at 50 ab$^{-1}$ Belle II with
gBG omitted.

\begin{figure*}[htbp]
	\begin{center}
		\includegraphics[width=0.45\textwidth]{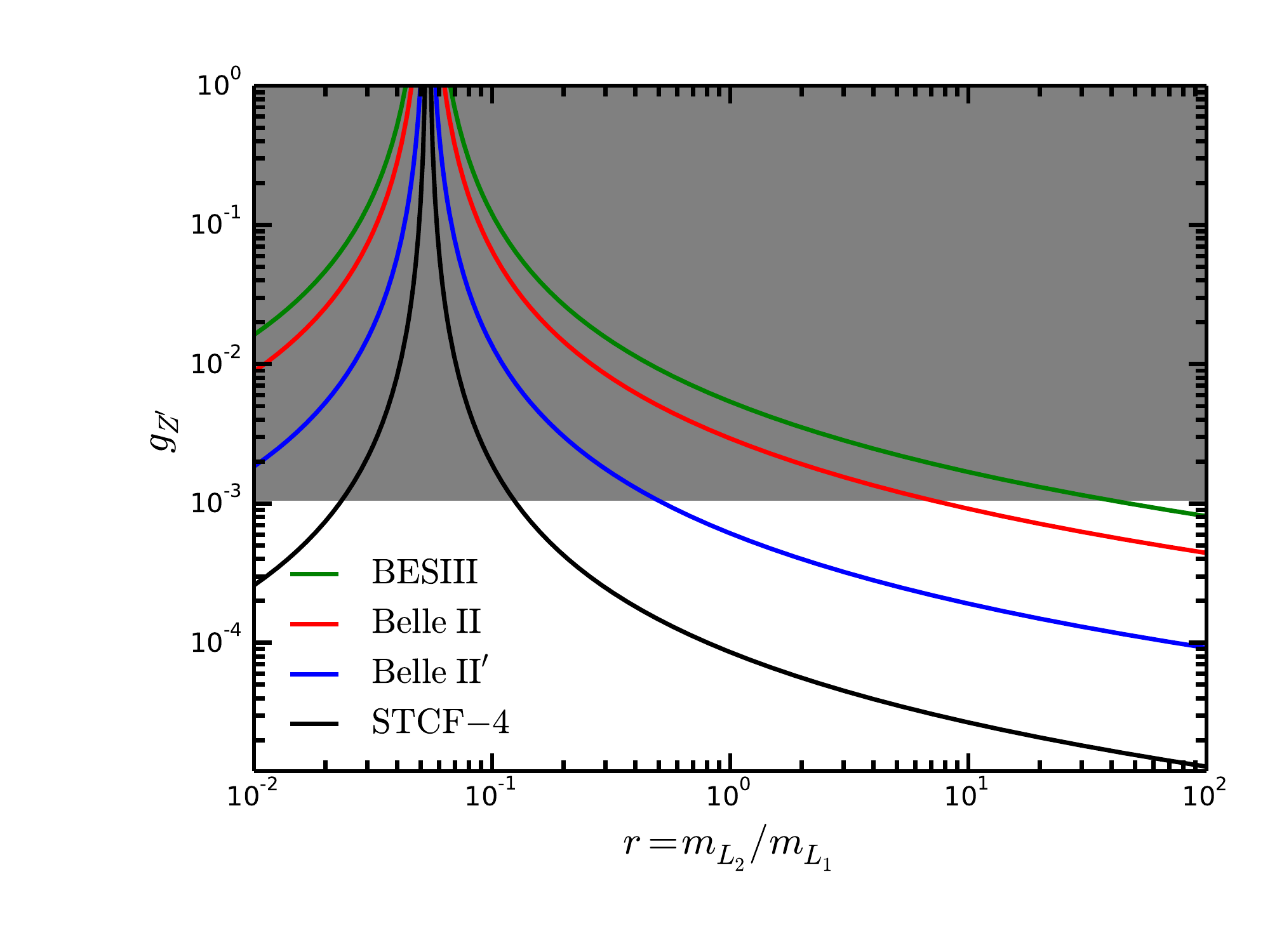}
		\includegraphics[width=0.45\textwidth]{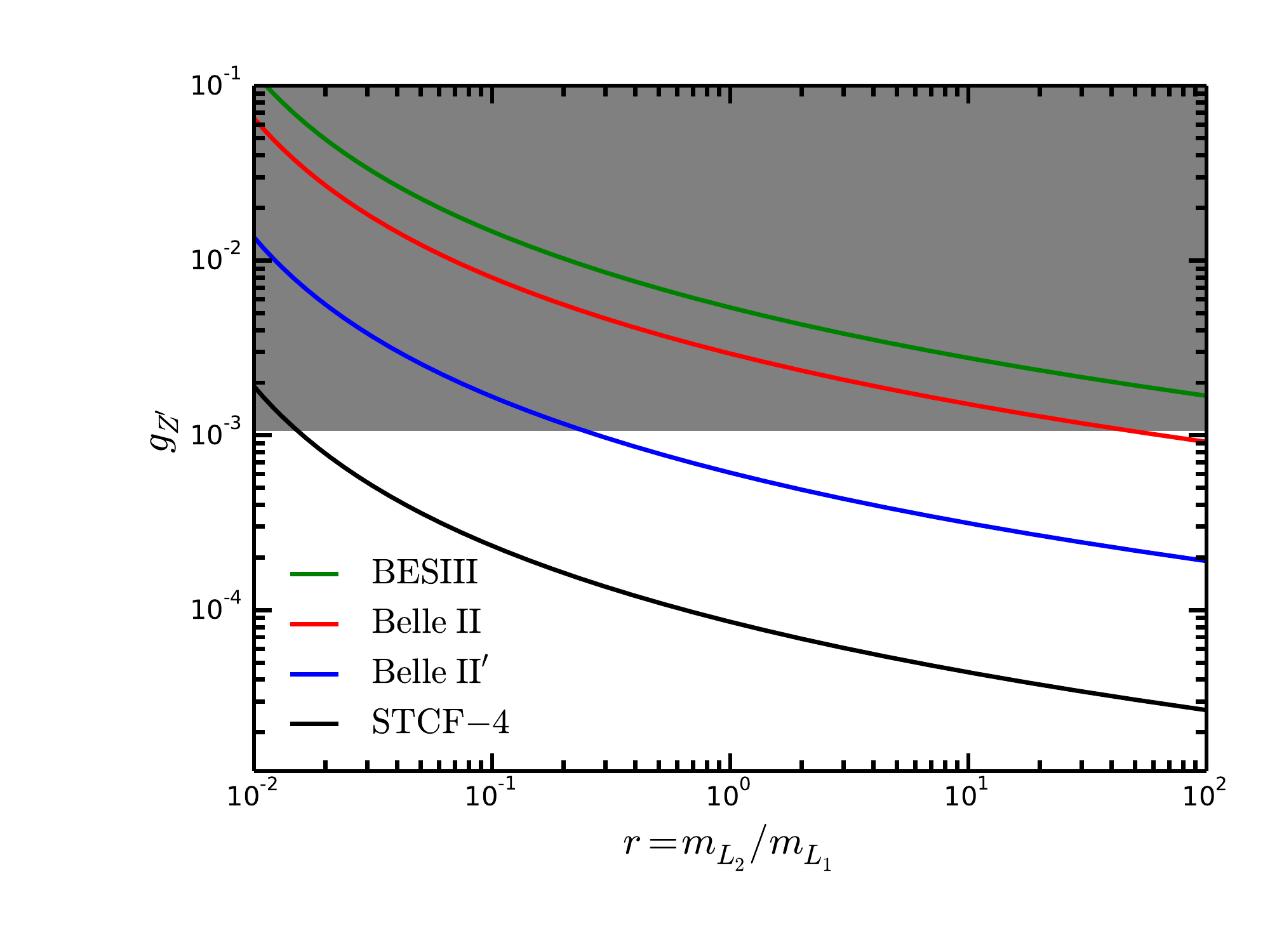}
		%\hspace{0.2cm}
		%\includegraphics[width=0.18\textwidth]{Figs/FeynDia-eevva}
		\caption{The expected 95\% C.L. exclusion limits in the $g_{\zp}-r$ plane for $\mzp=0.1$ GeV from BESIII with 14 fb$^{-1}$, Belle II with 50 ab$^{-1}$ and 4 GeV STCF
			with 30 ab$^{-1}$.
			The shaded grey region is already excluded by CCFR experiments, which is independent on the mass ratio.}
		\label{fig:gz-r}
	\end{center}
\end{figure*}

In Fig.\ref{fig:gz-r}, we present the dependence for exclusion regions of $g_{\zp}$ corresponding to  $\mzp=0.1$ GeV on the mass ratio $r=m_{L_2}/m_{L_1}$ and $r=m_{S_2}/m_{S_1}$ via 
monophoton searches from BESIII with 14 fb$^{-1}$, Belle II with 50 ab$^{-1}$ and future 4 GeV STCF
with 30 ab$^{-1}$.
The shaded grey region is already excluded by CCFR experiments, which is independent on the mass ratio.
One can see that $g_{\zp}$ can down to $1.3\ (2.7)\times10^{-5}$  when $m_{L_2}/m_{L_1}\ (m_{S_2}/m_{S_1})=100$ at 4 GeV STCF with 30 ab$^{-1}$.

%%%%% %%%%% %%%%% %%%%% %%%%%
%%%%% %%%%% %%%%% %%%%% %%%%%

\section{Summary} 
	
In this paper, we probe the invisible decay of the $\lmt$ gauge boson via monophoton signature
at three different electron colliders operated at the GeV scale: Belle II, BESIII, and STCF.
In the minimal $U(1)_{\lmt}$ model, we extend the SM with a  $U(1)_{\lmt}$ gauge symmetry and assume 
that the kinetic mixing term between $\zp$ and photon is absent at tree level, but can arise at one loop 
level due to $\mu$ and $\tau$ leptons. 	We also further extend the  minimal	$U(1)_{\lmt}$ model with 
extra heavy vector-like leptons or charged scalars, where the additional contributions to the 
kinetic mixing arising from extra particles inside the loop.
The exciting nondecoupling behavior of the contribution since the extra heavy vector-like leptons or 
charged scalars to the kinetic mixing is also demonstrated.
The visible signatures of  heavy leptons or charged scalars,
too heavy to be directly detected at high energy colliders,
maybe possible in processes modified by the $\gamma-\zp$ mixing.

We translate the sensitivity for dark photon within monophoton signature projected by Belle II to $U(1)_{\lmt}$ gauge boson taking into account various SM BGs.
We also recast the recent invisible search of $\zp$ in the $ \mu^+\mu^-\zp$ production at Belle II.
It is found that, 
By ignoring the BG due to the gaps in the detectors, we 
present the constraints at BESIII with 14 fb$^{-1}$ luminosity 
and at future 30 ab$^{-1}$ STCF.
For comparison, we also compute the limits at 50 ab$^{-1}$ Belle II without gBG taking into
account.
It is found that the future 2 GeV STCF can further improve the sensitivity to low mass $\zp$
than Belle II via monophoton signature since it is operated at lower energy.
The future STCF can exclude the moun $g-2$ anomaly favored parameter region when $\mzp \lesssim 5$ GeV.
And gauge coupling constant $g_{\zp}$ in the minimal $U(1)_{\lmt}$ model can be probed down to about $4.2\times 10^{-5}$ when $\mzp<2m_\mu$ at future 30 ab$^{-1}$ STCF with $\sqrt{s}=2$ GeV.

\section{Acknowledgement}
This work was supported in part by the National Natural Science Foundation of China (Grants No. 11805001, No.11935001).

\end{document}